\documentclass[authoryear,10pt]{elsarticle}

\usepackage{natbib}
\usepackage{epsfig}
\usepackage{graphicx}
\usepackage{bm}
\usepackage{amsmath,amssymb,amsthm,amsfonts,mathrsfs,textcomp,mathtools}
\usepackage{url}
\usepackage{multirow}
\usepackage{subfig}
\usepackage{aas_macros}
\usepackage[shortlabels]{enumitem}

\theoremstyle{definition}
\newtheorem{definition}{Definition}

\usepackage{hyperref,fullpage}
\newcommand{\clomon}{{\sc clomon}} 

\makeatletter
\def\namedlabel#1#2{\begingroup
    #2%
    \def\@currentlabel{#2}%
    \phantomsection\label{#1}\endgroup
}
\makeatother

\DeclareFontFamily{U}{mathx}{\hyphenchar\font45}
\DeclareFontShape{U}{mathx}{m}{n}{<-> mathx10}{}
\DeclareSymbolFont{mathx}{U}{mathx}{m}{n}
\DeclareMathAccent{\widebar}{0}{mathx}{"73}

\let\notORI\not 
\usepackage[mathb]{mathabx}
\let\not\notORI

\usepackage{xcolor} 
\definecolor{MBlu}{rgb}{.6,.6,3}
\definecolor{DBlu}{rgb}{.1,.1,.7}

\graphicspath{{figures/}}

\def\R{\mathbb{R}}

\usepackage{extarrows}
\usepackage[section]{placeins} 
\usepackage{booktabs}

\usepackage{tikz}
\usetikzlibrary{calc}
\usetikzlibrary{arrows}
\usetikzlibrary{fadings,decorations.pathreplacing}
\newcommand\pgfmathsinandcos[3]{
  \pgfmathsetmacro#1{sin(#3)}
  \pgfmathsetmacro#2{cos(#3)}
}

\newcommand\LatitudePlane[3][current plane]{
  \pgfmathsinandcos\sinEl\cosEl{#2} 
  \pgfmathsinandcos\sint\cost{#3} 
  \pgfmathsetmacro\yshift{\cosEl*\sint}
  \tikzset{#1/.style={cm={\cost,0,0,\cost*\sinEl,(0,\yshift)}}}
}

\newcommand\DrawLatitudeCircle[2][1]{
  \LatitudePlane{\angEl}{#2}
  \tikzset{current plane/.prefix style={scale=#1}}
  \pgfmathsetmacro\sinVis{sin(#2)/cos(#2)*sin(\angEl)/cos(\angEl)}
  \pgfmathsetmacro\angVis{asin(min(1,max(\sinVis,-1)))}
  \draw[current plane] (\angVis:1) arc (\angVis:-\angVis-180:1);
  \draw[current plane,dashed](180-\angVis:1) arc (180-\angVis:\angVis:1);
}
\tikzset{%
  >=latex, 
  inner sep=0pt,%
  outer sep=2pt,%
  mark coordinate/.style={inner sep=0pt,outer sep=0pt,minimum size=3pt,
    fill=black,circle}%
}

\journal{Celestial Mechanics and Dynamical Astronomy}
\begin{document}

\title{Use of the Semilinear Method to predict the Impact Corridor on
  Ground\thanks{This research was conducted under ESA contract
    No. 4000113555/15/DMRP “P2-NEO-II Improved NEO Data Processing
    Capabilities”.}
}

\author[sds]{L.~Dimare}
\ead{dimare@spacedys.com}
\author[sds,pi]{A.~Del~Vigna}
\ead{delvigna@spacedys.com}
\author[sds]{D.~Bracali~Cioci}
\author[sds]{F.~Bernardi}

\address[sds]{Space Dynamics Services s.r.l., via Mario Giuntini,
  Navacchio di Cascina, Pisa, Italy}
\address[pi]{Dipartimento di Matematica, Universit\`a di Pisa, Largo
  Bruno Pontecorvo 5, Pisa, Italy}

\begin{small}
    \begin{abstract}
        We propose an adaptation of the semilinear algorithm for the
        prediction of the impact corridor on ground of an
        Earth-impacting asteroid.  The proposed algorithm provides an
        efficient tool, able to reliably predict the impact regions at
        fixed altitudes above ground with 5 orders of magnitudes less
        computations than Monte Carlo approaches. Efficiency is
        crucial when dealing with imminent impactors, which are
        characterized by high impact probabilities and impact times
        very close to the times of discovery. The case of 2008 TC$_3$
        is a remarkable example, but there are also recent cases of
        imminent impactors, 2018 LA and 2019 MO, for which the method
        has been successfully used. Moreover, its good performances
        make the tool suitable also for the analysis of the impact
        regions on ground of objects with a more distant impact time,
        even of the order of many years, as confirmed by the test
        performed with the first batch of observations of Apophis,
        giving the possibility of an impact 25 years after its
        discovery.
    \end{abstract}
\end{small}
\maketitle

\begin{small}
    {\noindent\bf Keywords}: Near-Earth Asteroid, Impact Monitoring,
    Semilinear Method, Impact Corridor
\end{small}

\section{Introduction}
\label{intro}

The semilinear algorithm was first introduced in \cite{m1999} for the
recovery of lost asteroids. The aim was to have an efficient algorithm
to determine an approximation of the boundary of the recovery region
on the sky plane where the asteroid is supposed to be found at a given
time. This region is the image of the confidence region under the
strongly non-linear prediction map, making the linear approximation to
be not reliable.
To apply the prediction function directly on a sampling of the initial
confidence ellipsoid is not an efficient solution. Indeed, this is a
region in a six-dimensional space, hence a huge number of points is
needed to obtain a uniform sampling of it. On the other hand, many of
them correspond to very close observations by the prediction map.
With the semilinear method, non-linear effects are taken into account
by finding analytically a geometrical approximation of the
one-dimensional curve mapping to the confidence prediction
boundary. After that we need only to sample this curve to have a
representation of the boundary of the recovery region.

The semilinear technique is general and can be applied as well to a
map different from the prediction map. In \cite{mv1999} the method was
used for the impact risk assessment problem. In this work, the authors
used the semilinear projection of the uncertainty region on the
modified target plane, defined as the plane through the Earth's
centre, orthogonal to the asteroid velocity at the time of its closest
approach to the Earth, to perform the close approach analysis.

The semilinear algorithm reveals to be useful in order to reliably and
efficiently predict the impact region on ground for an asteroid with
high impact probability (IP), using far less computations than Monte
Carlo approaches.  Indeed, the semilinear method succeeds in providing
the boundary of the impact region on ground, with a number of
propagations 5 orders of magnitude less than Monte Carlo methods. The
efficiency of the algorithm is crucial when a potential impact is very
near in time, like for example in the case of an imminent impactor,
discovered only few days or even less than one day before impact, like
it happened for 2008 TC$_3$. In these cases, the algorithm proves to
be extremely efficient, giving the result in less than a minute,
without the need of parallelisation.

When applying the semilinear method to the prediction of the impact
region, we need to consider the impact map. Given a set of orbital
elements leading to an impact with the Earth, the impact map is
defined in a neighbourhood of it. If the impact is certain, the impact
region is defined by a projection of the propagated orbital
uncertainty over the two-dimensional surface of the Earth. In case
that the impact probability is not 1, we still have a connected set of
orbits compatible with the observations and leading to an impact at a
certain date, a so called virtual impactor (VI). In this case, the
impact region is obtained through the propagation of the intersection
of the orbital uncertainty region with the region of elements leading
to the impact.

The adaptation of the semilinear algorithm to the impact prediction
provides an efficient method to compute the boundary of the impact
region on ground. Furthermore, the target impact surface can be
defined at different altitudes on ground. The union of the boundaries
of the impact regions at different altitudes provides the impact
corridor, a curved tube inside which the asteroid falling trajectory
will lie with a certain likelihood. We can associate an impact
probability with each impact region, giving a measure of the
confidence that the real asteroid trajectory will actually be inside
the corridor.

This paper is organised as follows.
%
%
In Section~\ref{sec:semilinear} we describe the semilinear algorithm.
In Section~\ref{sec:ic_computation} we give the rigorous definition
of the impact map and provide the details of the algorithm for the
impact corridor computation.
In Section~\ref{sec:vi} we describe how an orbit is selected as VI
representative by the \clomon-2 impact monitoring system and how the
impact probability associated with the VI is computed.
In Section~\ref{sec:ip} we provide the equations for the computation
of the impact probabilities associated with the semilinear impact
regions.
In Section~\ref{sec:tests} we show the results of some meaningful
numerical tests, performed using real observational data.
Finally, in Section~\ref{sec:conclusions} we summarise the results of
this work and suggest some improvements.

\section{Outline of the semilinear method}
\label{sec:semilinear}

Let us consider the space $\R^N$ of the orbital elements, a
\emph{target space} $\mathcal{Y}\subseteq \R^2$, and a non-linear
function $F:W\subset\R^N\rightarrow \mathcal{Y}$, defined on an open
subset $W$ of $\R^N$. As usual, either $N=6$, if we consider a set of
six orbital elements (in whatever coordinates), or $N>6$ if some
dynamical parameter is included \cite[Chapter 1]{ODbook}, as it is the
case for impact predictions involving the Yarkovsky effect (see, for
instance, \cite{delvigna:yarko}). We assume that $F$ is continuously
differentiable on $W$, that is $F \in \mathcal{C}^1(W)$. Let us call
$\mathbf{x}$ and $\mathbf{y}$ the variables in the spaces $\R^N$ and
$\mathcal{Y}$, respectively.

Suppose to have a set of measurements and to solve for an orbit using
a least squares procedure, namely the differential corrections as
described in \cite[Section 5.2]{ODbook}. Let $\mathbf{x}_0\in \R^N$ be
the nominal solution of the least squares problem (the least squares
orbit), $\Gamma_X \in \R^{N\times N}$ the associated covariance matrix
($\Gamma_X=C_X^{-1}$, with $C_X$ the normal matrix) and let
$\mathbf{y}_0 = F(\mathbf{x}_0)$ be the nominal prediction.
The linear confidence ellipsoid $Z_{lin}^X(\sigma) \subset \R^N$
associated with the solution $\mathbf{x}_0$ is defined to be
\[
  Z_{lin}^X(\sigma) = \left\{\mathbf{x} \in \R^N \,:\,
  (\mathbf{x}-\mathbf{x}_0)^T C_X (\mathbf{x}-\mathbf{x}_0) \leq
  \sigma^2 \right\}.
\]
According to Gauss \cite{gauss1809}, the solution of a linear least
squares problem can be represented by a Gaussian probability density
function, with mean equal to the nominal solution and covariance
matrix equal to the inverse of the normal matrix, \emph{i.e.} the
matrix solving the normal equation of the differential correction last
step (that is, the one computed at convergence of the method).  In the
linear approximation, the boundaries of the linear confidence
ellipsoids represent the level curves of the Gaussian distribution of
the solutions.
%

Let $X$ denote the $N$-dimensional Gaussian random variable in the
initial space $\R^N$, with mean $\mathbf{x}_0$ and covariance
$\Gamma_X$. In the linear approximation, we consider the variable
$Y=DF_{\mathbf{x}_0}(X)$, which is the image of $X$ through the
differential of $F$ at $\mathbf{x}_0$. As it is known from the theory
of Gaussian probability distributions, the random variable $Y$ is also
Gaussian, with covariance matrix given by
\[
  \Gamma_Y = DF_{\mathbf{x}_0} \Gamma_X DF_{\mathbf{x}_0}^T.
\]
The confidence ellipsoid $Z_{lin}^X(\sigma)$ is mapped onto an
elliptic disk in the target space, which we denote by
$Z_{lin}^Y(\sigma)$, defined by the inequality
\[
  (\mathbf{y}-\mathbf{y}_0)^T C_Y (\mathbf{y}-\mathbf{y}_0)\leq
  \sigma^2,
\]
where $C_Y=\Gamma_Y^{-1}$ is the normal matrix of $Y$. Assuming that
the differential $DF_{\mathbf{x}_0}$ has rank 2, and that the matrix
$C_X$ is non-degenerate, the matrix $C_Y$ is non-degenerate too and
the disk $Z_{lin}^Y(\sigma)$ is a two-dimensional surface with an
ellipse $\mathcal{E}_Y(\sigma)$ as boundary.
The boundary ellipse $\mathcal{E}_Y(\sigma)$ is the image through
$DF_{\mathbf{x}_0}$ of an ellipse $\mathcal{E}_X(\sigma)$ in the
orbital elements space, which lies on the boundary of the ellipsoid
$Z_{lin}^X(\sigma)$. We define the \emph{semilinear confidence
  boundary} $K(\sigma)$ as the non-linear image in the target space of
the ellipse $\mathcal{E}_X(\sigma)$, that is
\[
  K(\sigma) = F(\mathcal{E}_X(\sigma)).
\]
By the Jordan curve theorem \cite[pp. 587-594]{jordan1887}, if the
closed curve $K(\sigma)$ has no self-intersection points, then it is
the boundary of a connected subset $Z(\sigma)$ in $\mathcal{Y}$. The
subset $Z(\sigma)$ is the semilinear approximation of
$F(Z_{lin}^X(\sigma))$.

To compute the semilinear confidence boundary $K(\sigma)$ we can
proceed as follows. The rows of the Jacobian matrix
$DF_{\mathbf{x}_0}$ span a 2-dimensional subspace $\mathcal{G}$ in the
orbital elements space $\R^N$, which can be decomposed as
\[
  \R^N = \mathcal{G}\oplus \mathcal{H},
\]
where $\mathcal{H} = \mathcal{G}^\perp$ is a
$(N-2)$-dimensional subspace\footnote{Without any further indication,
  we mean that the orthogonal subspace is taken with respect to the
  Euclidean scalar product in $\R^N$.}. We can define a rotation
matrix $R\in \R^{N\times N}$ such that
\[
  R\left(\mathbf{x}-\mathbf{x}_0\right) =
  \begin{pmatrix}
    \mathbf{g}-\mathbf{g}_0 \\ \mathbf{h}-\mathbf{h}_0
  \end{pmatrix},
\]
where the vector $\mathbf{g}$ represents two coordinates in the space
$\mathcal{G}$ and $\mathbf{h}$ represents $N-2$ coordinates in the
orthogonal space.
In the new coordinate system the normal matrix $RC_XR^T$
can be decomposed as
\[
RC_XR^{T} =
\begin{pmatrix}
C_{\mathbf{gg}} & C_{\mathbf{gh}} \\
C_{\mathbf{hg}} & C_{\mathbf{hh}}
\end{pmatrix}.
\]
The equation
\[
\mathbf{h}-\mathbf{h}_0=
-C_{\mathbf{hh}}^{-1}C_{\mathbf{hg}}(\mathbf{g}-\mathbf{g}_0)
\]
defines a 2-dimensional subspace in $\R^N$, containing the points of
the confidence ellipsoid $Z_{lin}^X(\sigma)$ with tangent space
orthogonal to $\mathcal{G}$. This is called regression subspace of
$\mathbf{h}$ given $\mathbf{g}$ \cite[Section 5.4]{ODbook}, and we
denote it as $\mathcal{R}$. The space $\mathcal{G}$ can be mapped to
the regression subspace $\mathcal{R}$ by means of the map $H$, defined
as follows
\begin{equation}\label{eq:regr}
  \begin{array}{cccc}
    H:&
    \mathcal{G}
    &\longrightarrow
    &\mathcal{R}
    \\
    &\mathbf{g}-\mathbf{g}_0 & \mapsto &
    \begin{pmatrix}
    \mathbf{g}-\mathbf{g}_0\\
    -C_{\mathbf{hh}}^{-1}C_{\mathbf{hg}}(\mathbf{g}-\mathbf{g}_0)
    \end{pmatrix}.
    \end{array}
\end{equation}
The map above defines a parametrisation of $\mathcal{R}$ with
coordinates $\mathbf{g}$. The image through
$DF_{\mathbf{x}_0}$ of $\mathcal{G}$ gives the image of the
entire $N$-dimensional space $\R^N$. Since $DF_{\mathbf{x}_0}$
is assumed of rank 2, it can be described as
\begin{equation}
  \label{eq:df_split}
  DF_{\mathbf{x}_0}=A\circ \Pi_{\mathbf{g}}\circ R,
\end{equation}
where $\Pi_{\mathbf{g}}$ is the orthogonal projection on $\mathcal{G}$
and $A: \mathcal{G} \rightarrow \mathcal{Y}$ is an invertible
$2\times2$ matrix.  Using equation \eqref{eq:df_split} we obtain (see
\cite[Section 7.5]{ODbook})
\[
  \Gamma_Y=A\Pi_{\mathbf{g}}R\Gamma_XR^T\Pi_{\mathbf{g}}^T A^T
  =A\Gamma_{\mathbf{gg}}A^T,
\]
where $\Gamma_{\mathbf{gg}}$ is the marginal covariance matrix in the
space $\mathcal{G}$.
Then $A^{-1}(\mathcal{E}_Y(\sigma))$ is an ellipse in $\mathcal{G}$,
and the ellipse $\mathcal{E}_X(\sigma)$ on the boundary surface of the
ellipsoid is its image under the map $H$ defined by
\eqref{eq:regr}. In other words, the ellipse $\mathcal{E}_X(\sigma)$,
mapping to the semilinear boundary, corresponds to the intersection of
the boundary surface of the ellipsoid $Z_{lin}^X(\sigma)$ with the
regression subspace $\mathcal{R}$.

As already pointed out in the introduction, the advantage in using the
semilinear approximation of the non-linear image
$F(Z_{lin}^X(\sigma))$ comes from the representation of its boundary
through a one-dimensional curve in the space $\R^N$ of initial
conditions. This means that we are able to numerically represent the
region $F(Z_{lin}^X(\sigma))$ through the sampling of a 1-dimensional
curve instead of the entire $N$-dimensional region
$Z_{lin}^X(\sigma)$. Thus we need $r^{N-1}$ less points to obtain the
same resolution $r$.

\section{Impact Corridor Computation}
\label{sec:ic_computation}

We developed an adaptation of the semilinear algorithm recalled in
Section~\ref{sec:semilinear} for the prediction of the impact corridor
on ground for an asteroid that has a non-zero chance of impacting the
Earth in the future. In this Section we provide the details of the
implemented algorithm. Possible improvements and extention of
functionalities are described in Sections~\ref{sec:vi} and
\ref{sec:ip}.

It is worth pointing out that the semilinear method is an approximation.
First of all, the linear approximation of the uncertainty region in
the initial orbital elements space is used. Second, the method
involves a linearization of the map from the initial orbital elements
space to the target space.  This linearization is used to select
properly a representative curve in the initial orbital elements space,
which is non-linearly propagated, taking into account all the relevant
perturbations, to obtain the boundary of the predicted uncertainty on
the target space. It follows that the non-linear effects are
considered only in part. As a consequence, the approximation is more
reliable when the uncertainty is small, or otherwhise in the vicinity
of the nominal solution. In the original method, this corresponds to
the vicinity to the point where the prediction map is linearized,
which is in turn another condition for reliability. In the adaptation
of the method for the purpose of the impact corridor computation these
two conditions in general do not coincide, because the linearization
is performed at a point that does not necessarily coincide with the
nominal solution (see Section~\ref{sec:ic_steps}).

Another important aspect is the computational load, which is directly
connected to the IP value. In particular, as it will be clear later in
Section~\ref{sec:optimisation}, the computational load is higher for
lower values of the $IP$. This, together with the fact that this kind
of computation is less interesting when the $IP$ is very low, is the
reason why we decided to impose a threshold $IP_{\rm th}>0$ to the
impact probability, under which the method is not applied.
We considered $IP_{\rm th}=1\cdot 10^{-3}$.

We have implemented the
algorithm within the OrbFit
5.0\footnote{\url{http://adams.dm.unipi.it/orbfit/}} software used by
the online information systems NEODyS and AstDyS. Consequently, the
integration of the equations of motion is performed using the 15-th
order version of RADAU integrator \cite{ra15}. This code is not
distributed.

In the following paragraph, we recall some basic concepts and
terminology of orbit determination and impact monitoring theory, which
are needed to understand the algorithm for the impact corridor and the
context where it is used. This introduction is not intended to explain
or even introduce the general theory of impact monitoring, as it would
be out the scope of the paper. It contains only what is strictly
necessary to clearly explain the presented work.

The starting point of our algorithm for the impact corridor
computation is a least squares orbit of an asteroid with impact
probability $IP>IP_{\rm th}$. In general, to have a positive $IP$ does
not imply that the nominal solution impacts the Earth. The least
squares solution comes with an uncertainty, which defines the region
in the orbital elements space where the real orbit can be with a
certain level of confidence, as provided by a probability distribution
defined on the basis of the covariance matrix \cite[Chapter
5]{ODbook}. To have a positive, not negligible, impact probability
means that the orbits corresponding to a part of the uncertainty
region impact the Earth. This subset of the uncertainty region does
not necessarily contain the nominal solution, unless the probability
is almost 1.

In summary, since $IP>0$, there exists a set of orbits leading to an
impact and still compatible with the observations.  This implies that
there exists a \emph{Virtual Impactor (VI)}, which is defined as a
connected set of initial conditions leading to an impact at about the
same date and compatible with the least squares solution
\cite{milani:visearch}.  In order to locate the impact region on
ground we need to conveniently select a \emph{VI Representative}, a
specific point of the VI. Using the least squares solution and the VI
representative, the semilinear method provides the boundary of the
impact region at a selected altitude above ground, roughly
corresponding to the portion of the initial uncertainty region that
leads to the impact.

Another basic concept to be taken into account in order to understand
the overall context of this work, is the \emph{Target Plane
  (TP)}. Given the planetocentric position and velocity of the
asteroid nominal solution at the time $\bar{t}$ of minimum distance
from the Earth, the TP
is defined to be the plane passing through the Earth's centre of mass
and orthogonal to the incoming asymptote of the hyperbola defining the
two-body approximation of the trajectory at the time $\bar{t}$ of
closest approach \cite{valsecchi:resret}. The TP is not used
  directly by the semilinear method, but it is a fundamental tool in
  impact monitoring (IM), the output of which is the starting point of
  the impact corridor computation.  The TP
  is used in IM for the return analysis. In particular, the impact
  probability associated with a given VI is computed using a suitably
  defined Gaussian probability density function on the TP, which is
  integrated over the Earth impact cross section, as described in
  \cite{mcstv2005} and recalled in Section~\ref{sec:ip_vi}. Then, an
  explicit impacting orbit is searched for by looking inside the Earth
  impact cross section on the TP and this is the selected VI
  representative.

In this work, besides using part of the IM output directly as input
for the semilinear method, the basic theory behind the $IP$
computation will be exploited to define an impact probability
associated with the impact region on ground, as will be explained in
Section~\ref{sec:ip}. Moreover, in
Section~\ref{sec:vi_representative}, we will suggest how to select a
VI representative, which is most suitable as input to the impact
corridor algorithm in order to obtain the most reliable result.

\subsection{Preliminary definitions}
\label{sec:ic_defs}

In order to give a precise definition of impact at a certain altitude
above ground, usable in numerical tests, we consider the Earth's
geometric reference ellipsoidal surface, defined by the WGS 84 model
\cite{wgs84}. In this approximation the Earth's surface is a
geocentric ellipsoid of revolution with semimajor axis equal to
$6378.137$~km and flattening parameter $f$ defined by the equality
$1/f=298.257223563$. The eccentricity $e$ can be derived from the
relation $e^2=f(2-f)$.

\begin{definition}
For a fixed altitude $h>0$, the \emph{impact surface} $S_h$ at
altitude $h$ above ground is the surface of points in space at
altitude $h$ above the WGS 84 Earth reference ellipsoid, which is in
turn the impact surface $S_0$ at zero altitude\footnote{Note that only
  $S_0$ is an ellipsoid, whereas $S_h$ is not an ellipsoid, for any
  $h>0$.}.
\end{definition}

\begin{definition}
Let $\sigma>0$ and $h\geq 0$ be fixed values for the confidence
parameter and the altitude. Given a VI, the corresponding \emph{impact
  region boundary} $\mathcal{B}_{\sigma,h}$ at altitude $h$ above
ground and confidence level $\sigma$ is the result of the propagation
of the intersection of the VI with the boundary of the uncertainty
ellipsoid $Z_{lin}^X(\sigma)$ corresponding to the selected $\sigma$,
until the impact surface at altitude $h$ above ground is reached.
\end{definition}

\begin{definition}
The \emph{impact corridor} $\mathcal{C}_\sigma$ corresponding to the
confidence level $\sigma$ is the union of the boundaries of the impact
regions from the nominal altitude of atmospheric entry $h_{\rm max}=100$~km, to
the ground. In symbols
\[
    \mathcal{C}_\sigma = \bigcup_{0\leq h\leq h_{\rm max}}
    \mathcal{B}_{\sigma,h}.
  \]
\end{definition}

\subsection{Inputs and Impact Map}
\label{sec:impact_map}

Let $\mathbf{x}_0\in \R^N$ be the nominal orbit and $\Gamma_X$ its
covariance matrix, both provided at some epoch $t_0$. Since the
nominal orbit may not impact, what matters is the VI representative
orbit. Let $\mathbf{x}_{imp}\in \R^N$ be the orbit of the VI
representative, provided at the same initial epoch $t_0$ as the
nominal orbit.  Both the full least squares solution
$(\mathbf{x}_0,\Gamma_X)$ and the VI representative orbit
$\mathbf{x}_{imp}$ at time $t_0$ must be provided as input to the
procedure.

For a fixed altitude $h$, with $0 \leq h \leq h_{\rm max}$, the
impact map $F^h$ to the surface $S_h$ at altitude $h$ above ground is
defined in a neighbourhood $W$ of $\mathbf{x}_{imp}$:
\[
F^h: W \subset \R^N \rightarrow S_h.
\]
In order to explicitly compute the result of the mapping to the impact
surface, we have to perform three steps. First, we need to
compute the time $t^*$ at which the orbit reaches the surface
$S_h$. Then we propagate the state vector to this time. Finally, we
project the propagated state vector to the corresponding position on
$S_h$.
The map $F^h$ is thus defined as the following composition:
\begin{equation}
  \label{eq:impact_map}
\begin{array}{cccccccc}
  F^h: W \subset
&\R^N
&\xlongrightarrow{\Psi}& \R^{N+1}
&\xlongrightarrow{\Phi}  & \R^N
&\xlongrightarrow{\Sigma_h}& S_h
\\
&\mathbf{x}
&\mapsto & (t^*,\mathbf{x})
&\mapsto & \mathbf{z}^*=\mathbf{z}(t^*,\mathbf{x})
&\mapsto & \mathbf{y}
\end{array}
\end{equation}
In the above formula, the vector $\mathbf{x}\in W$ is the initial
orbit at time $t_0$ and the three maps are as follows:
\begin{enumerate}[label={\upshape(\arabic*)},wide = 0pt,leftmargin=*]
  \item the map $\Psi$ is obtained through the computation of the
    impact time $t^*$ with the surface $S_h$;
  \item the map $\Phi$ is the integral flow solving the equations of
    motion and giving the state vector
    $\mathbf{z}=(\mathbf{p},\mathbf{v})$, with $\mathbf{p}$ and
    $\mathbf{v}$ the heliocentric Cartesian position and velocity. It
    is evaluated at the impact time $t^*$, so that
    $\mathbf{z}=\mathbf{z}^*$;
  \item the map $\Sigma_h$ is the projection of the state vector
    $\mathbf{z}$ to the surface $S_h$, giving the impact position
    $\mathbf{y}=F^h(\mathbf{x})\in S_h$ in geodetic coordinates. Let
    $(\lambda,\varphi,\zeta)$ be the geodetic coordinates on the WGS
    84 ellipsoid, being $\lambda$ the longitude, $\varphi$ the
    latitude and $\zeta$ the altitude, so that
    $\mathbf{y}=(\lambda,\varphi)$.
\end{enumerate}
We are assuming to solve the equations of motion using heliocentric
Cartesian coordinates. Indeed this is the case with the OrbFit
software. In what follows we assume that the state vector $\mathbf{z}$
is given in the J2000 equatorial reference frame.

We denote by $\tau$ the map that gives the impact time $t^*$ as
function of the initial orbit $\mathbf{x}\in \R^N$, that is
\[
t^*=\tau(\mathbf{x}).
\]
The map $\tau$ is implicitly defined, imposing that the altitude is
$h$:
\begin{equation}
\label{eq:tau_map}
\zeta(\mathbf{p}(t,\mathbf{x}))-h=0 .
\end{equation}
The algorithm to compute the impact time $t^*$ uses the classical
\emph{regula falsi} method (see, for instance, \cite{ConteDeboor1972})
applied to the function $f(t) = \zeta(\mathbf{p}(t, \mathbf{x}))-h$,
defining the distance to the impact surface $S_h$. The check for the
occurrence of an impact and the consequent computation of the impact
time $t^*$ have to be done when the object is experiencing a close
approach with the Earth, that is when the geocentric distance of the
object is less than $d_\oplus=0.2$~au, as it is common for impact
monitoring purposes.

Let $t_k$ and $t_{k+1}$ be two consecutive steps of the
propagation of the orbit inside the close approach interval.
The sign of $f(t_k)$ and $f(t_{k+1})$ is checked at any
step $k$ to establish the occurrence of
an impact. Let the integer parameter $\delta$ assume value $1$ in case
that the propagation is forward in time, value $-1$ in case that it is
backward. If
\[
f(t_k)\delta>0 \quad \text{and} \quad f(t_{k+1})\delta<0,
\]
then the sign of $f$ changes in a way that is compatible with the
orbit approaching and impacting the surface $S_h$ in the interval
$(t_k,t_{k+1})$. In this case the \emph{regula falsi} is applied to
the function $f$ to find the time $t^*\in (t_k,t_{k+1})$, where it
becomes zero. The current software determines the intersection with
the impact surface with a precision of $10^{-3}$~km.

\subsection{Steps of the algorithm}
\label{sec:ic_steps}

The application of the semilinear method consists in following the
steps described in Section~\ref{sec:semilinear}, using the impact map
$F^h$. The main adaptation comes from the fact that the function $F^h$
in general is not defined on the entire curve $\mathcal{E}_X(\sigma)$.
We can summarise the steps for the impact corridor computation as follows.
\begin{enumerate}[label={\bf IC\arabic*},wide = 0pt,leftmargin=*]
\item \label{step1} The confidence region $Z_{lin}^X(\sigma)$ is
  linearly propagated using the differential of $F^h$ at
  $\mathbf{x}_{imp}$. This allows us to obtain the linear confidence
  region $Z_{lin}^Y(\sigma)$ on the tangent space to $S_h$ in
  $F^h(\mathbf{x}_{imp})$.
\item \label{step2} The linear approximation given by
  $DF^h_{\mathbf{x}_{imp}}$ is exploited to select a representative
  curve $\mathcal{E}_X(\sigma)$ on the boundary of the initial
  confidence ellipsoid $Z_{lin}^X(\sigma)$, in fact an ellipse.
\item \label{step3} A finite sampling of the ellipse
  $\mathcal{E}_X(\sigma)$ is propagated non-linearly, including all
  the relevant perturbations, to obtain the predicted semilinear
  boundary at altitude $h$. For this step we have to distinguish
  between two cases.\vspace{0.05cm}
  \begin{enumerate}[label={\bf IC3\alph*},wide = 0pt,leftmargin=*]
  \item \label{step3a} For $IP = 1$ all the points of the sample
    are propagated. It is possible to launch many processes in
    parallel to decrease the processing time when the sample is large.
  \item \label{step3b} For $IP < 1$ there are two possibilities:
    \begin{itemize}
    \item an optimisation procedure is applied, avoiding the
      propagation of many non-impacting points contained in the
      sample.
    \item alternatively, many processes can be launched in parallel in
      order to decrease the processing time while propagating all the
      orbits of the sample.
    \end{itemize}
    It is not possible to combine the
sation procedure actually
    implemented with parallel computing.
  \end{enumerate}
\end{enumerate}

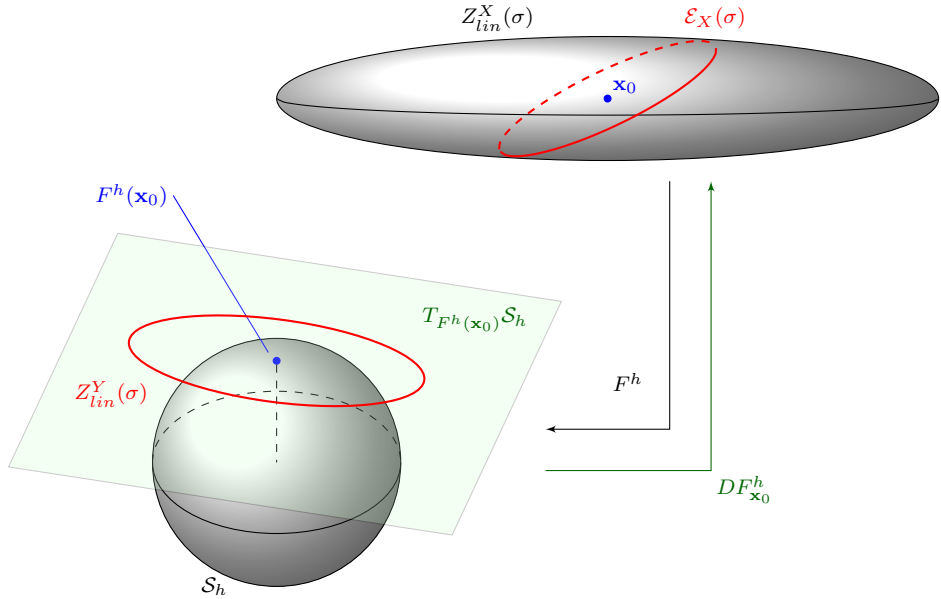
\begin{figure}[!ht]
  \centering
  \begin{tikzpicture}[scale=1.1]
    \def\R{1.5}      
    \def\angEl{35}   
    \def\angAz{-105} 
    \def\angPhi{-40} 
    \def\angBeta{19} 

    \def\a{4}
    \def\b{0.75}
    \def\bb{0.2}
    \def\ag{\b*sqrt((4*\a*\a+1-3*\b*\b)/(\a*\a+1))}
    \def\bg{\ag/16}

    \coordinate (O) at (0,0);
    \coordinate (N) at (1,3.5);
    \coordinate (V) at ($(3/4*\a,\b*1.2)+(N)$);

    \coordinate[mark coordinate, blue] (V) at ($(3/4*\a,\b*1.2)+(N)$);
    \node[anchor=south west] at (V) {\footnotesize{\color{blue}
        {$\mathbf{x}_0$}}};

    \draw [ball color=white] (V) ellipse (\a cm and \b cm);
    \draw[smooth,samples=100,domain=-\a:\a,thin]
    plot(\x+3/4*\a+1,{-\bb/\a*sqrt(\a*\a-\x*\x)+\b*1.2+3.5});
    \node at ($(-\a/3,\b*1.3)+(V)$) {\footnotesize{$Z_{lin}^X(\sigma)$}};
    \coordinate[mark coordinate, blue] (V) at ($(3/4*\a,\b*1.2)+(N)$);
    \node[anchor=south west] at (V) {\footnotesize{\color{blue}
        {$\mathbf{x}_0$}}};

    \draw[-latex'] ($(V)+(0.75,-1)$) -- ($(V)+(0.75,-1)+(0,-3)$)
    -- ($(V)+(0.75,-1)+(0,-3)+(-1.5,0)$);%

    \node[anchor=north west] at ($(V)+(0,-3.25)$)
    {\footnotesize{$F^h$}};%

    \pgfmathsetmacro\H{\R*cos(\angEl)}
    \tikzset{xyplane/.style={
        cm={cos(\angAz),sin(\angAz)*sin(\angEl),
          -sin(\angAz),cos(\angAz)*sin(\angEl),(0,-\H)} } }
    \LatitudePlane[equator]{\angEl}{0}
    \coordinate (P) at (0,\H);

    \fill[ball color=white] (0,0) circle (\R);
    \draw (0,0) circle [line width=0mm,radius=\R];
    \node at (-0.5*\R,-\R) {\footnotesize{$\mathcal{S}_h$}};

    \coordinate[mark coordinate,blue] (P) at (0,\H);
    \node[anchor=east] at ($(P)+(-1.25,2)$)
    {\color{blue}\footnotesize{$F^h(\mathbf{x}_{0})$}};
    \draw[blue,thin] ($(P)+(-0.1,0.1)$) -- ($(P)+(-1.25,2)$);

    \DrawLatitudeCircle[\R]{0}
    \draw[dashed] (P) -- (O);

    \filldraw[xyplane,shift={(P)},fill=green!20,opacity=0.2]
    (-1.4*\R,-1.7*\R) rectangle (2*\R,2*\R);
    \node[green!40!black] at (1.6*\R,1.15*\R)
    {\footnotesize{$T_{F^h(\mathbf{x}_{0})}\mathcal{S}_h$}};

    \node[anchor=north west] at ($(V)+(1.25,-1)+(0,-3.5)$)
         {\color{green!40!black}
      \footnotesize{$DF^h_{\mathbf{x}_{0}}$}};%

    \draw[rotate=-7.5, red, thick] (P) ellipse (1.8 cm and 0.5 cm);
    \node at ($(P)+(-2,-0.4)$)
    {\color{red}\footnotesize{$Z_{lin}^Y(\sigma)$}};

    \draw[smooth,samples=100,
    domain=-\ag*0.999:\ag*0.999,thick,red,rotate around={25.7:(V)}]
    plot(\x+3/4*\a+1,{-\bg/\ag*sqrt(\ag*\ag-\x*\x)+\b*1.2+3.5});

    \draw[smooth,samples=100,
    domain=-\ag*0.999:\ag*0.999,thick,red,rotate around={25.7:(V)},dashed]
    plot(\x+3/4*\a+1,{\bg/\ag*sqrt(\ag*\ag-\x*\x)+\b*1.2+3.5});

    \node at (\a/5+3/4*\a+1.5,\b*1.3+\b*1.2+3.5)
    {\color{red}{\footnotesize{$\mathcal{E}_X(\sigma)$}}};

    \draw[latex'-,green!40!black] ($(V)+(1.25,-1)$) -- ($(V)+(1.25,-1)+(0,-3.5)$)
    -- ($(V)+(1.25,-1)+(0,-3.5)+(-2,0)$);%
\end{tikzpicture}
  \caption{Graphical sketch of the application of the semilinear
  method described above in the steps \ref{step1}-\ref{step3}, for the
  case with IP$ = 1$. In this case $\mathbf{x}_{imp}=\mathbf{x}_{0}$.}
  \label{fig:semil_ic}
\end{figure}

By applying the above steps for $h=100$~km and $h=0$~km, we obtain a
semilinear representation of the impact corridor corresponding to the
confidence level $\sigma$.
Steps from \ref{step1} to \ref{step3} are graphically represented in
Figure~\ref{fig:semil_ic} for the case with $IP = 1$.

For the general case, with $IP < 1$, we exploit the linear
approximation of $F^h$ in a neighbourhood of $\mathbf{x}_{imp}$,
giving
\[
F^h(\mathbf{x})-F^h(\mathbf{x}_{imp})\simeq
DF^h_{\mathbf{x}_{imp}}\left(\mathbf{x}-\mathbf{x}_{imp} \right).
\]
The linear map defines an isomorphism between the ellipse
$\mathcal{E}_X(\sigma)$ centred in the nominal solution $\mathbf{x}_0$
and an ellipse in the tangent plane to $S_h$ in
$F^h(\mathbf{x}_{imp})$, with centre
$DF^h_{\mathbf{x}_{imp}}(\mathbf{x}_0-\mathbf{x}_{imp})$, since
\[
DF^h_{\mathbf{x}_{imp}}\left(\mathbf{x}-\mathbf{x}_{imp} \right)=
DF^h_{\mathbf{x}_{imp}}\left(\mathbf{x}-\mathbf{x}_0 \right)+
DF^h_{\mathbf{x}_{imp}}\left(\mathbf{x}_0-\mathbf{x}_{imp} \right).
\]
We have to consider the intersection of the ellipse
$\mathcal{E}_X(\sigma)$ with the open connected set $W$ of impacting
orbits (the interior of the VI), in order to obtain the set of points
where the map $F^h$ is defined. In other words, the impact
map is defined for the points near $\mathbf{x}_{imp}$, that in general
may be far from the nominal solution, if the IP is small.

As a final remark, we observe that it is assumed that the map
$DF^h_{\mathbf{x}_{imp}}$ has rank $2$, otherwise the algorithm cannot
be applied. This is a reasonable assumption, as it is generally
satisfied by real orbits. In case that a degeneracy occurs, a
possibility is to propagate the orbit together with its uncertainty,
and also the selected VI representative, to a time $t_0'\neq t_0$ near
to $t_0$ and start the impact corridor algorithm, using the propagated
solution and VI representative as input.

\subsection{Differential of the Impact Map}
\label{sec:derivatives}
In order to perform the steps outlined in Section~\ref{sec:ic_steps}
we need to compute the differential $DF^h$ of the impact map at the VI
representative initial orbit $\mathbf{x}_{imp}\in \R^N$.
Using the notations introduced in Section~\ref{sec:impact_map},
we have
\[
F^h=\Sigma_h\circ \Phi \circ \Psi = G \circ \Psi ,
\]
where
\[
G(t,\mathbf{x})=\left(\Sigma_h\circ \Phi\right) (t,\mathbf{x})
= (\lambda(\mathbf{p}(t,\mathbf{x})),\varphi(\mathbf{p}(t,\mathbf{x})))
\quad \textrm{and} \quad
\Psi(\mathbf{x})=(\tau(\mathbf{x}),\mathbf{x}) .
\]
Then the differential at $\mathbf{x}_{imp}$ is
\[
DF^h_{\mathbf{x}_{imp}} = \frac{\partial G}{\partial t}(t^*,\mathbf{x}_{imp})
\frac{\partial \tau}{\partial \mathbf{x}}(\mathbf{x}_{imp}) +
\frac{\partial G}{\partial \mathbf{x}}(t^*,\mathbf{x}_{imp}) .
\]
Since the map $\Sigma_h$ depends only on the position, we have
\[
\frac{\partial G}{\partial t} (t^*,\mathbf{x}_{imp})=
\frac{\partial \Sigma_h}{\partial \mathbf{p}} (\mathbf{p}^*)
\frac{\partial  \mathbf{p}}{\partial t} (t^*,\mathbf{x}_{imp})=
\frac{\partial (\lambda,\varphi)}{\partial \mathbf{p}} (\mathbf{p}^*)
~\mathbf{v}^* ,
\]
where $\mathbf{p}^*$ and $\mathbf{v}^*$ are the heliocentric position
and velocity at the impact time $t^*$.
From equality \eqref{eq:tau_map} the derivative of $\tau$ is, by the
implicit function theorem,
\[
\frac{\partial \tau}{\partial \mathbf{x}}(\mathbf{x}_{imp}) = -
\left(\frac{\partial \zeta} {\partial t}
  (t^*,\mathbf{x}_{imp})\right)^{-1}
\frac{\partial \zeta} {\partial
  \mathbf{x}} (t^*,\mathbf{x}_{imp}) ,
\]
with
\[
\frac{\partial \zeta} {\partial t} (t^*,\mathbf{x}_{imp}) =
\frac{\partial \zeta} {\partial \mathbf{p}} (\mathbf{p}^*)
\frac{\partial \mathbf{p}} {\partial t} (t^*,\mathbf{x}_{imp})=
\frac{\partial \zeta} {\partial \mathbf{p}} (\mathbf{p}^*) \mathbf{v}^*
\]
and \[
\frac{\partial \zeta} {\partial \mathbf{x}} (t^*,\mathbf{x}_{imp}) =
\frac{\partial \zeta} {\partial \mathbf{p}} (\mathbf{p}^*)
\frac{\partial \mathbf{p}} {\partial \mathbf{x}} (t^*,\mathbf{x}_{imp}) .
\]
The term $\partial G/ \partial \mathbf{x} (t^*,\mathbf{x}_{imp})$ is
\[
\frac{\partial G}{\partial \mathbf{x}}(t^*,\mathbf{x}_{imp}) =
\frac{\partial \Sigma_h}{\partial \mathbf{p}} (\mathbf{p}^*)
\frac{\partial \mathbf{p}}{\partial \mathbf{x}} (t^*,\mathbf{x}_{imp}) =
\frac{\partial (\lambda,\varphi)}{\partial \mathbf{p}} (\mathbf{p}^*)
\frac{\partial \mathbf{p}}{\partial \mathbf{x}} (t^*,\mathbf{x}_{imp}) .
\]
In the above equations the term $\partial \mathbf{p}/ \partial
\mathbf{x} (t^*,\mathbf{x}_{imp})$ comes from the integration of the
variational equation. The geodetic coordinates and their derivatives
are computed from the explicit expression of the geocentric equatorial
position as function of them.

\subsection{Optimisation}
\label{sec:optimisation}

As previously pointed out, for virtual impactors with $IP = 1$ all the
points of the curve $\mathcal{E}_X(\sigma)$ lead to an impact. Thus to
obtain a satisfactory sample of the semilinear boundary it suffices to
sample the curve with a few hundred points. On the contrary, for a
virtual impactor with $0< IP <1$, in general the points of the ellipse
$\mathcal{E}_X(\sigma)$ do not necessarily impact and the subset of
impacting points is usually not enough to obtain a clear
representation of the semilinear boundary, even with the possibility
to obtain no impacting points at all. Indeed the fraction of impacting
points among the sampling is roughly proportional to $IP$, thus impact
probabilities of the order of $0.001$ require about $100,000$ points
to obtain a proper visualisation of the impact corridor.

Such a high number of orbits to propagate leads in turn to very long
computational times, so that an optimisation procedure is needed to
propagate the least possible number of non-impacting orbits, since
they do not contribute to the semilinear boundary sample. Different
procedures can be implemented by exploiting the symmetry of the
ellipse $\mathcal{E}_X(\sigma)$ with respect to its semimajor axis
(the projection of the weak direction on the regression space).  If
the stretching, which is the size of the tangent vector to the Line Of
Variation (LOV) trace on the TP \cite{mcstv2005}, is high, the
confidence ellipse on the TP is very elongated and we can assume an
approximated symmetry of the intersection of $\mathcal{E}_X(\sigma)$
with the impacting region. This assumption is not reliable for very
low values of the stretching. Anyway, when the stretching is low the
impact probability turns out to be high, so that the impact corridor
can be readily obtained through the non-optimised procedure, that is
by propagating all the sample orbits.

Let us assume that the stretching is high and that
$\mathcal{E}_X(\sigma)\cap W$ consists of two disjoint symmetric
segments.  This is generically true when $\sigma$ is greater than the
value associated to the VI representative $\mathbf{x}_{imp}$,
otherwise the above intersection may be a single segment or
empty. With simple numerical tricks the implemented procedure is able
to manage all these cases correctly, without the need to know a priori
which is the case at hand.
The optimisation procedure works as follows. We first search for two
reference impacting orbits among the points of the sampling of the
ellipse $\mathcal{E}_X(\sigma)$, one for each of the two impacting
segments.
Let $i_1$ and $i_2$ be the indices in the sampling of the two found
impacting orbits. In general, they fall in the middle of the two
impacting segments and then the algorithm search for the four
endpoints of the two segments. To this end, four loops of propagations
are performed. Starting from $i_1$, a forward and a backward loop
propagate consecutive sample orbits until the first non-impacting
orbit is found. The same is done starting from $i_2$. In particular,
the orbits with indices $i_1,\,i_1+1,\,\dots$ and
$i_1-1,\,i_1-2,\,\ldots$ are propagated until the first non-impacting
orbit is found, and the same is done starting from the index $i_2$.

To find the two impacting orbits, one for each impacting segment, the
algorithm proceeds as follows. We divide the ellipse in four equal
parts, delimited by its axes, and we perform a procedure which is
similar to a binary search in each quarter of ellipse.  As a
preliminary step, we check whether the endpoints of the quarter are
impacting orbits. If none of them impacts, we consider the middle
point of the quarter. If it does not impact as well, we have two
consecutive segments of the quarter of the ellipse, each delimited by
two non-impacting orbits. We now apply the same procedure to both
segments and iterate. At any subsequent step, we consider the middle
points of the intervals obtained in the previous step.

If we have propagated $n$ points without finding an impact, in the
next step we consider $n-1$ new points. Each one is the mean
of two consecutive points tried in the previous step.
As a result, at the end of the $k$-th step, with $k\geq 0$, we have
propagated at most $2^k+1$ points (step $k=0$ corresponds to the
propagation of the two extremes).  Let $N_{q}$ be the total number of
sample orbits of the considered quarter of ellipse. If $N_{q}=2^k+1$
we need exactly $k$ steps to perform the check for impact on all the
$N_{q}$ points. If $2^k+1 < N_{q} < 2^{k+1}+1$, we must perform $k+1$
steps to check all the $N_{q}$ points.

The procedure stops as soon as an impacting orbit is found, returning
the index of the orbit in the original sampling. The modified binary
search is repeated for each quarter of ellipse in an ordered way, and
the search stops as soon as a total of two impacting orbits are
found. The order selected for the scan of the quarters is based on the
value of the parameter $\sigma_{\rm LOV}$ corresponding to the
location of the VI representative $\mathbf{x}_{imp}$ along the LOV.
The symmetry assumption is also considered.  In this way, we have that
most of the times it suffices to perform the search only for two
quadrants. This is the reason why we chose to divide the ellipse in
four parts instead of considering the two semi-ellipses symmetric with
respect to the semimajor axis.

The optimisation procedure may fail if there are two VIs with near
impact times (\emph{e.g.}, same day of impact). If we are in this
case, there are two possibilities.
\begin{itemize}
  \item The VI chosen for the impact corridor computation and the
    other one have values of $\sigma_{\rm LOV}$ with opposite sign. In
    this case the procedure succeeds in finding the impact corridor of
    the selected VI.
  \item The VI chosen and the other one have $\sigma_{\rm LOV}$ with
    the same sign. In this case the procedure is not able to
    distinguish between the two VIs and may result in the propagation
    of the points of the non-selected VI. Then, the impacting segments
    returned at the end of the impact corridor computation may be the
    ones of the non-selected VI. Even worse, the two impacting
    segments in output may belong to two different VIs, so that the
    resulting boundary does not make sense. These cases should be
    treated by deactivating the optimisation and propagating all the
    sample points. The output will contain the impact region
    boundaries coming from both VIs, not only from the selected one.
\end{itemize}
Most of the times the occurrence of two VIs with close impact times in
the risk table is misleading, and the two reported VIs are actually
segments of the same connected set. This is a drawback of the
algorithms used in the impact monitoring process. For the purpose of
IM, to have more representatives of the same VI does not cause
concern. These cases are treated well by the optimisation procedure,
that is able to find the entire impacting region whatever the selected
VI representative.

\section{Virtual Impactor Characterisation}
\label{sec:vi}

The virtual impactors and the associated impact probabilities are the
main outcome of the impact monitoring process. Two impact monitoring
systems are currently operating worldwide, \clomon-2 and Sentry,
developed respectively at the University of Pisa and at NASA-JPL. They
are extensively described in \cite{mcstv2005}. Both systems provide a
characterisation of all the virtual impactors found for a specified
asteroid through the risk table, a list of virtual impactors along
with their characterising parameters: the impact date, the impact
probability, the value of the sigma parameter corresponding to the VI
location along the LOV.

We use as starting point the results provided by the \clomon-2 impact
monitoring system of the NEODyS-2 public
service\footnote{\url{https://newton.spacedys.com/neodys/}}, which is
the system that our group at SpaceDyS is in charge of maintaining and
managing for ESA.  In particular, the algorithm for the impact
corridor computation needs as input the asteroid orbit provided with
its uncertainty and a VI representative, \emph{i.e.} a set of orbital
elements that belongs to the VI and leads to an impact with the Earth.

In what follows we recall how the \clomon-2 system selects a VI
representative and how it computes the impact probability associated
with the VI.

\subsection{Selection of the VI representative}
\label{sec:vi_representative}

To characterise the VI, we need to select properly a VI
representative, which is an explicitly computed set of initial
conditions belonging to the VI. The VI representative
is given as output by the \clomon-2 system, for each computed VI. As
it is evident from the description of the algorithm for the impact
corridor computation provided in Section~\ref{sec:ic_computation}, in
order to obtain the best approximation of the impact region, we need
to select a point which best approximate the centre of the VI.

The representative currently provided by \clomon-2 is not always
guaranteed to satisfy this condition, since for impact monitoring
purposes this is not needed. Usually the solutions provided by the
sampling of the LOV do not impact and the system searches for an
impacting orbit through different iterative procedures, by looking for
the minimum of the distance from the Earth's centre of mass. The
system stops the search for the minimum if it finds an impacting orbit
and this particular solution is selected as VI representative. In the
case that the sampling of the LOV is dense enough to give many orbits
of the same VI, the software selects the representative that leads to
the minimum distance from the Earth centre of mass.
In this last case, if the LOV geometry is not too complicated and
consequently the interval of the LOV curve crossing the Earth's
section is well approximated by a chord (no high curvature, no curls),
and if the sampling of the LOV is dense enough at the VI location, the
selected representative is actually near the VI centre.

The \clomon-2 system functionalities are currently being migrated to
the software of the NEOCC\footnote{\url{https://neo.ssa.esa.int}}. The
migration includes some improvements of both the impact monitoring
algorithms and software. A major improvement is the possibility to
densify the LOV sampling, when a return consists of a few points
\cite{delvigna:densification}.  Given a denser sampling of the
intersection between the VI and the LOV, the selection of the VI
representative is automatically improved. Apart from highly stretched
uncertainty ellipsoids with negligible width, it is still not
guaranteed that the representative is close to the VI centre. For high
curvature, it is not even guaranteed that it is close to the centre of
the LOV impacting segment.

In order to obtain a representative near the VI centre, we need a new
procedure, able to perform two different computations, depending on
the value of the width parameter. In case of negligible width, the
procedure should connect the sigma parameter with the distance along
the LOV in order to catch the centre of the LOV impacting segment. In
the case that the width is not negligible, the procedure should be
able to find the centre of the VI off LOV.


For the test results described in Section~\ref{sec:tests}, we did not
change the VI representative selection of the \clomon-2 system and we
used the output as currently provided by it.

\subsection{Impact Probability associated with the VI}
\label{sec:ip_vi}

A fundamental output of the impact monitoring system is the estimation
of the impact probability associated with the found VI. The \clomon-2
system performs a sampling of the LOV, returning a set of Virtual
Asteroids (VAs) along the LOV, each with its covariance matrix.  For
any VA, the system performs a non-linear analysis of the close approach
on the TP. Then, as described in \cite{mcstv2005}, a probability density
function is defined on the TP. It is derived from the Gaussian
probability density defined on the orbital elements space on the basis
of the least square solution uncertainties of the nominal orbit and of
the VAs. The impact probability associated with the VI is obtained
integrating the TP probability density over the cross-sectional area
of the Earth.

In the linear approximation, the uncertainty ellipse associated to
each VA maps to an uncertainty ellipse on the TP, centred on the TP
trace of the VA nominal orbit. The semimajor and semiminor axes of the
ellipse on the TP are the stretching $s$ and the semiwidth
$w$. A Gaussian probability density $p_{\rm TP}$ is defined on the TP
as the product of two univariate probability densities with variances
$s$ and $w$. Given that the VA is not actually the nominal
orbit, a correction is applied to the univariate probability density
defined along the direction of the semimajor axis.

We take the coordinates $(u,v)$ on the TP defined along the directions
of the semiminor and semimajor axis, with origin in the centre of the
Earth. If $(u',v')$ are the coordinates of the VI representative on
the TP and $\sigma_{\rm LOV}$ is the LOV parameter corresponding to
the selected VI representative, we have
\[
p_{\rm TP}(u,v)=p_1(u)p_2(v) ,
\]
where
\[
p_1(u) = \frac{1}{{\sqrt{2\pi} w}}\exp\left( - \frac{1}{2} {\left(
  {\frac{{u - u' }}{w}} \right)^2 } \right) , \quad
p_2(v) = \frac{1}{{\sqrt{2\pi} s}}\exp\left( - \frac{1}{2}
{\left( {\frac{{v - v' }}{s} + \sigma_{\rm LOV} } \right)^2
} \right) .
\]

Then the impact probability is
\[
IP = \int\!\!\!\int_{\mathcal{S}_\oplus} p_{\rm TP}(u,v)\;dv\;du\; ,
\]
where $\mathcal{S}_\oplus$ is the Earth impact cross section on the
TP. In numerical computations, the above integral is restricted to the
domain of TP points with distance from the LOV trace less than $8w$.

The \clomon-2 system computes the above integral only when the VI
representative is on the LOV. The system is able to detect also
off-LOV VIs, with an explicit nominal impacting orbit out of the
LOV. In this case, a correction factor $k_{\rm off}$ is applied, instead
of recomputing the probability integral. The factor $k_{\rm off}$ is
defined as
\[
k_{\rm off}=e^{-\frac{1}{2}(\chi^2-\sigma_{\rm LOV}^2-\sigma_{\rm imp}^2)},
\]
where $\chi$ is, up to a factor depending on the number of
observations, the RMS of the residuals, $\sigma_{\rm LOV}$ is the LOV
parameter corresponding to the VI and $\sigma_{\rm imp}$ is the
lateral distance from the LOV to the Earth impact cross section
divided by the semi-width $w$.

Finally, the \clomon-2 system uses a 1-dimensional approximation,
when the piece of the LOV intersecting the Earth's cross section on the
TP is long enough and the sampling of the LOV is dense enough to give
more than 10 impacting VAs. Let $\{\sigma_i\}_{i=1,\,\ldots,\,N}$ be
the values of the LOV parameter corresponding to the target plane
points inside the Earth impact cross section. The IP is computed as
\[
IP = \frac{1}{\sqrt{2\pi}} \sum_{i=1}^{N-1}
e^{-\frac{\sigma_i^2}{2}}(\sigma_{i+1}-\sigma_{i}) ,
\]
which is the approximate value of the integral
\[
\int_{\sigma_1}^{\sigma_N} \frac{1}{\sqrt{2\pi}}e^{-\frac{\sigma^2}{2}}d\sigma_{\rm LOV}
\]
with the rectangle method.

\section{Impact Probability associated with the Corridor}
\label{sec:ip}

The impact corridor computed with the semilinear method is obtained
using the restriction of the impact map $F^h$ to the regression
subspace $\mathcal{R}\subset \R^N$, defined in Section
\ref{sec:semilinear}. We take this into account in order to properly
associate an impact probability to the impact regions obtained with
this method.  To this end, we define a suitable probability density
function $p_R$ on the regression subspace $\mathcal{R}$.  Even the
impact probability associated to the VI can be estimated using the
probability distribution defined on $\mathcal{R}$ and considering the
points of the VI belonging to $\mathcal{R}$. This gives a value $IP_R$
of the IP associated with the selected VI congruent with the method,
so that the semilinear projection of the entire VI on ground will have
an associated $IP$ equal to $IP_R$. The level of approximation
of the IP computed in this way is directly connected to the
approximation coming from the semilinear projection. The value $IP_R$
in general differs from the value returned by the IM system and
described in Section~\ref{sec:ip_vi}. A conditional probability is
also defined through the density $p_R$, imposing the impact event
related to the selected VI.

The impact map $F^h$ is defined on the entire VI. Its restriction to
the regression subspace $\mathcal{R}$ is obtained considering the set
$V=(\Pi_{\mathbf{g}}\circ R)(VI\cap \mathcal{R})$, which is the
intersection of the VI with the regression subspace, projected on the
plane $\mathcal{G}$. Then we define the impact map $G^h$ on $V$, using
the lifting $H$ to the regression space defined in Section
\ref{sec:semilinear}:
\[
\begin{array}{cccc}
G^h:&V\subset \mathcal{G} &\longrightarrow& S_h \\
& \mathbf{g} &\mapsto& (F^h \circ R^T \circ H) (\mathbf{g}) .
\end{array} \]
From the definition of $G^h$, it is evident that the following
relation holds
\[F^h|_{VI\cap \mathcal{R}}=G^h\circ \Pi_{\mathbf{g}}\circ R .\]
The image through $F^h$ of the intersection $VI\cap \mathcal{R}$ is
the semilinear prediction of the impact region corresponding to the
entire VI. Indeed, the semilinear approximation consists in selecting
the regression subspace $\mathcal{R}$ on the basis of the differential
of $F^h$ at the VI representative.

According to \cite{gauss1809}, the least squares solution
$\mathbf{x}_0$  is the mean of a
Gaussian probability distribution on the initial conditions space
$\R^N$, whose density function is given by
\[
p_X(\mathbf{x})=N(\mathbf{x}_0,\Gamma_X)=\frac{\sqrt{\det C_X}}
{(2\pi)^{N/2}}\exp\left(-\frac{1}{2}(\mathbf{x}-\mathbf{x}_0)^T
C_X (\mathbf{x}-\mathbf{x}_0)\right) ,
\]
where $\Gamma_X$ is the covariance matrix of the least square solution
and $C_X = \Gamma_X^{-1}$.  We define $p_R$ to be the marginal
probability density of $X$ on $\mathcal{G}$, given by
\[
  p_R(\mathbf{g})=N\left((\Pi_{\mathbf{g}}\circ
  R)(\mathbf{x}_0),(\Pi_{\mathbf{g}}\circ R)
  \Gamma_X(\Pi_{\mathbf{g}}\circ R)^T\right) .
\]
For any subset $A\subset \mathcal{G}$ we define the associated
probability
\[ P_R(A)=\int_{A}{p_R(\mathbf{g})d\mathbf{g}}\ .\]
The IP associated with the VI through $p_R$ is
\[IP_{R}=\int_V{p_R(\mathbf{g})d\mathbf{g}} .\]
In order to take into account that we are considering the intersection
of the VI with the regression subspace $\mathcal{R}$, and not the
projection of the entire VI on $\mathcal{G}$, we have to correct the
probability density $p_R$ with a scaling factor $k$, based on the IP
associated to the entire VI. We use the value of the IP computed by
the IM system, as described in Section~\ref{sec:ip_vi}, so that
\[\tilde{p}_R=kp_R , \quad \textrm{with} \quad k=IP/IP_R .\]
We use the density $\tilde{p}_R$ to infer the probability $P_I(B)$ that the
asteroid impacts on a region $B\subset S_h$. We take the counter image
$A=(G^h)^{-1}(B)=\{\mathbf{g}\in \mathcal{G} : G^h(\mathbf{g}) \in B\}$, so that
\[
P_I(B)=kP_R(A) .
\]

Finally, for any subset $A\subset \mathcal{G}$, we define a
conditional probability, forcing the impact corresponding to the
selected VI:
\[P_R(A|V)=\frac{\int_{A\cap V}{p_R(\mathbf{g})d\mathbf{g}}}{IP_{R}}\ .\]
If $A=(G^h)^{-1}(B)$, with $B\subset S_h$, this gives the probability
that the asteroid impact location is inside the region $B\subset S_h$,
given the assumption that the impact foreseen by the selected VI
happens.

\section{Numerical tests}
\label{sec:tests}

We tested our method on five asteroids, namely 2008~TC$_3$, (99942)
Apophis, 2014~AA, 2018~LA and 2019~MO.
We compared the results obtained for 2008~TC$_3$ and Apophis with
that computed by an independent system and with a different method,
that is the JPL impact regions computed with a Monte Carlo
simulation. For both the objects we obtained a remarkable agreement.
The semilinear method succeeds in providing the boundary of the impact
region on ground, with a comparatively smaller number of propagations
with respect to Monte Carlo approaches. Indeed it samples a
1-dimensional curve instead of a region in the 6-dimensional orbital
elements space.

In the case of 2014~AA, the non-linear propagation causes a twist of
the uncertainty region, so that the plotted boundary does not give
enough information to identify the extension of the impact region. In
other words, the impact region expands outside the twisted boundary.

For the cases of the recent imminent impactors 2018~LA and 2019~MO,
the semilinear predictions are in good agreement with the fireball
detections, even though there are very few observations.

Concerning the graphical representation of the impact corridor, the
output of the semilinear procedure is a data file with geodetic
coordinates representing points on the Earth surface. It is then
needed to plot them on the terrestrial globe and we exploited the
already existing software Google Earth for the figures of this
section.

For all the tests we used OrbFit version 5.0, extended with the module
for the computation of the semilinear impact regions.  Unless
otherwise specified, we used the following options for the astrometric
error model and the dynamical model:
\begin{itemize}
\item automatic biases removal and weighting, using the scheme
  described in \cite{fcct2015};
\item use of JPL ephemerides DE431 \cite{Folkner_etal_2014} for the
  Newtonian terms of the Sun, the eight planets, Pluto, and the Moon;
\item inclusion of the perturbations from 16 massive main-belt
  asteroids, whose ephemerides are computed with OrbFit. They are
  Ceres, Pallas, Juno, Vesta, Hebe, Iris, Hygiea, Eunomia, Psyche,
  Amphitrite, Europa, Cybele, Sylvia, Thisbe, Davida, Interamnia;
\item inclusion of the relativistic terms for the eight planets,
  according to the Einstein-Infeld-Hoffmann model
  \cite[Sec. 4]{Moyer2003};
\item addition of the effect of the Earth oblateness in the vicinity
  of the Earth (distance less than 0.1 au), with $J_2=0.0010826267$.
\end{itemize}
We performed the impact monitoring using the following options:
\begin{itemize}
\item non-linear Line Of Variations, obtained through constrained
  differential corrections, as described in \cite{mstac2005};
\item uniform sampling in probability of the solutions along the LOV,
  as described in \cite{delvigna2018,delvigna_etal2019};
\item sampling for the uncertainty parameter $\sigma_{\rm LOV}$ in the
  interval $[-5,5]$.
\end{itemize}

\subsection{Prediction of the impact corridor for 2008~TC$_3$}
\label{sec:2008tc3}
Asteroid 2008~TC$_3$ was discovered by Richard A. Kowalski at the
Catalina Sky Survey on October 6$^{\rm th}$, 2008 at 6:39 UTC. The
object entered the atmosphere above the Nubian Desert in northern
Sudan on October 7$^{\rm th}$, 2008 at 2:46 UTC, just 20 hours after
its first detection. It has been the first body to be observed and
tracked prior to fall on the Earth.
At the time of the first detection, 2008~TC$_3$ was more distant from
Earth than the Moon. It was soon recognised as a possible impactor
with probability practically 1, so that many astronomers put their
efforts in observing it and now we have available an observational
dataset composed by nearly 900 observations. Furthermore, the asteroid
actual ground track is known well, thanks to the meticulous work of
recovery of many meteorites that reached the desert floor by the
University of Khartoum \cite{shaddad_etal_2010}. The availability of
many observations and the knowledge of the ground track give us the
opportunity to validate our impact location software against real
data.

We performed two tests. First we analysed the results of the orbit
determination and the impact location prediction, obtained with the
full dataset of 883 observations available before impact. We compared
our results with those obtained by NASA-JPL. As a second test, we
analysed the evolution of the impact corridor prediction, using
different reduced observational data sets. This second test was aimed
at giving an indication on when it is useful to predict the impact
corridor on ground.

The JPL team provides a precise estimate of the trajectory of
2008~TC$_3$ and its impact ground track in \cite{fjrcdc2017}.  They
performed the orbit determination of 2008~TC$_3$ after a careful
analysis of the astrometric dataset and the selection of the weights
to assign to each observation. From one side they accounted for the
expected quality of some observers, and on the other side they
deweighted the observations toward the end of the arc since they show
a gradually poorer quality.
We adopted the same scheme for weights and outlier rejections as the
one used in \cite{fjrcdc2017}, where the JPL solution
18\footnote{\url{https://ssd.jpl.nasa.gov/sbdb.cgi?sstr=2008TC3;cad=1}}
is considered\footnote{The weights and the rejections used by the JPL
  were communicated to us by Davide Farnocchia, through a private
  communication.}. Consequently 308 observations were rejected as
outliers.
Moreover, we employed the same high-precision force model, which is
the one adopted in all the tests of this paper, described at the
beginning of Section \ref{sec:tests}.
We applied the semilinear method using directly the impacting nominal
solution as VI representative.

\begin{figure}[!ht]
  \centering
  \includegraphics[width=0.6\textwidth]{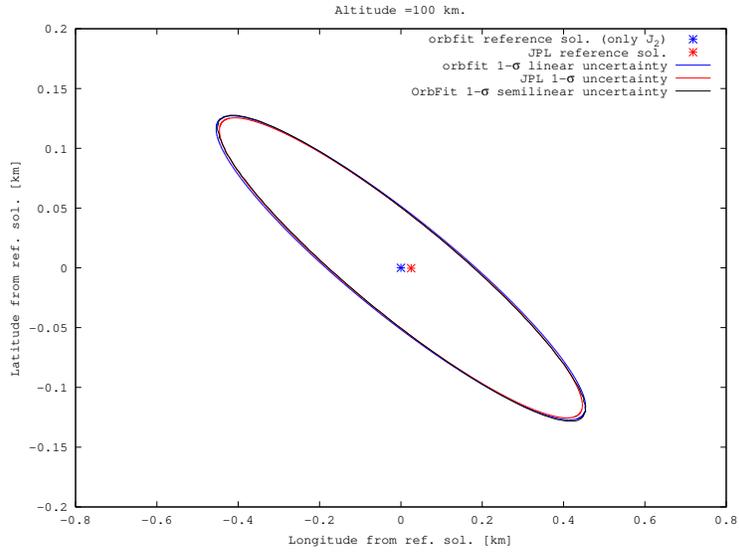}
  \caption{Nominal impact location and 1-$\sigma$ uncertainty region
    at 100 km altitude for 2008~TC$_3$. The linear uncertainty
    predictions by JPL and OrbFit, corresponding to the parameters of
    Table~\ref{tab:comparison_jpl-orbfit}, are compared with the
    semilinear prediction. }
  \label{fig:linear_unc}
\end{figure}

The outcomes of the impact location prediction on ground were
compared (see Table~\ref{tab:comparison_jpl-orbfit}). As reported in
\cite{fjrcdc2017}, we know that if we start the propagation from the
same nominal solution, the difference in the nominal impact location
due to the use of different numerical integrators is as large as
3~m. Recomputing the orbital solution, the difference slightly
increases.
\begin{table}
  \caption{Impact parameters and linear uncertainty at 100~km
    altitude, obtained by JPL and by OrbFit for 2008~TC$_3$.}
  \label{tab:comparison_jpl-orbfit}
  \centering
  \begin{tabular}{p{0.3\linewidth}p{0.3\linewidth}p{0.3\linewidth}}
    \toprule
    \textbf{Parameter} & \textbf{JPL Solution 18} &
    \textbf{OrbFit Solution}\\
    \toprule
    Time UTC & 2008-Oct-07 02:45:30.33 $\pm$ 0.14~s
    & 2008-Oct-07 02:45:30.31 $\pm$ 0.14~s\\ \midrule
    Latitude & 21.0871$^\circ \pm$ 0.0011$^\circ$ &
    21.0871$^\circ \pm$  0.0011$^\circ$ \\\midrule
    East Longitude &  30.5380$^\circ \pm$ 0.0043$^\circ$ &
    30.5378$^\circ \pm$ 0.0043$^\circ$\\\midrule
    1-$\sigma$ semimajor axis & 0.461~km &
    0.469~km \\\midrule
    1-$\sigma$ semiminor axis & 0.049~km &
    0.050~km \\\midrule
    Major axis azimuth & 104.6$^\circ$ & 104.5$^\circ$ \\\midrule
    1-$\sigma$ north-south uncertainty & 0.125~km & 0.127~km\\\midrule
    1-$\sigma$ east-west uncertainty & 0.446~km& 0.454~km\\
    \bottomrule
  \end{tabular}
\end{table}
This case is linear and the orbit is over-determined, so that the
1-$\sigma$ uncertainty region is very small, about $0.05\times 0.5$~km
at impact on ground. Figure~\ref{fig:linear_unc} shows the nominal
impact location and the uncertainty region at 100~km altitude. We
reported the two ellipses obtained by OrbFit and by JPL with the
linear approximation, corresponding to the parameters of
Table~\ref{tab:comparison_jpl-orbfit}. We also reported the semilinear
prediction. As we can see, the difference in the nominal impact
location is all in longitude and the three regions almost overlap,
being the OrbFit uncertainty slightly larger than the JPL one.

\begin{figure}[!ht]
  \centering
  \includegraphics[width=0.9\textwidth]{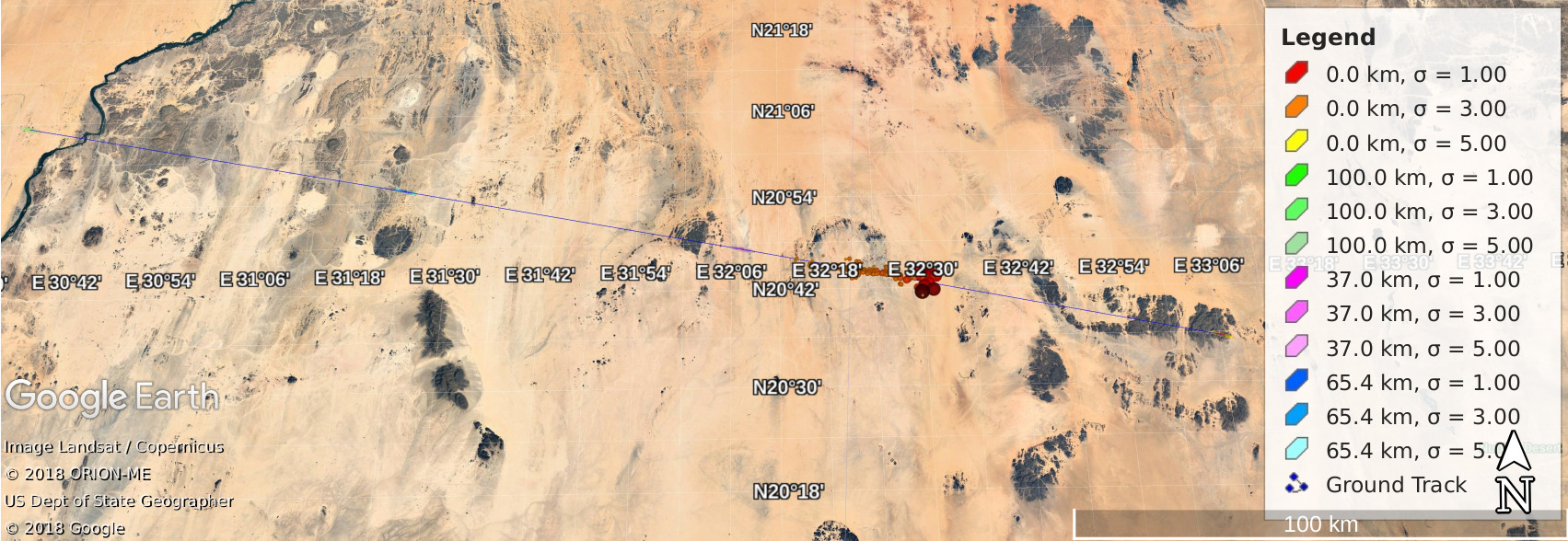}
  \caption{Prediction of the impact region of 2008~TC$_3$ with 883
  observations.}
  \label{fig:all_obs_TC3}
  \vspace{0.3cm}
  \includegraphics[width=0.9\textwidth]{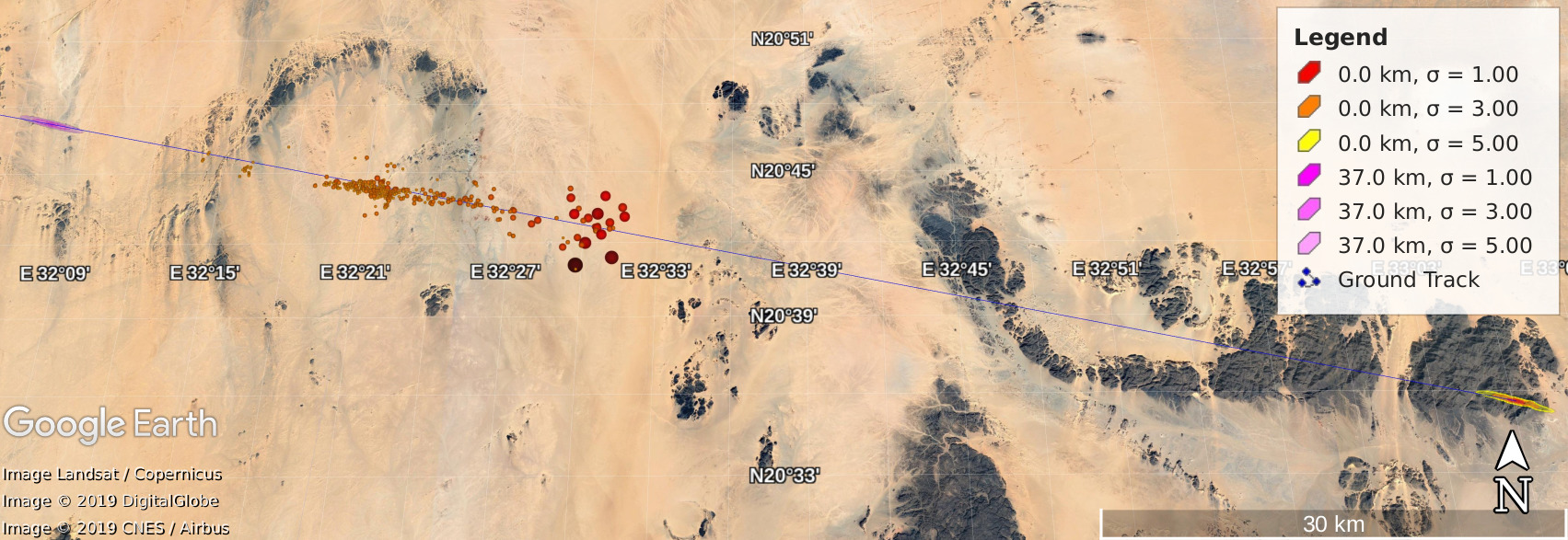}
  \caption{Enlargement of Figure \ref{fig:all_obs_TC3} between $h=37$~km and
  $h=0$~km.}
  \label{fig:all_obs_TC3_zoom}
\end{figure}

In Figure~\ref{fig:all_obs_TC3} we show the 2008~TC$_3$ impact regions
on ground and for altitudes corresponding to 37~km, 65.4~km and
100~km. Figure~\ref{fig:all_obs_TC3_zoom} is just an enlargement of
Figure~\ref{fig:all_obs_TC3} between $h=37$~km and
$h=0$~km. Detections of the actual atmospheric impact event suggested
an atmospheric entry at 65.4~km, followed by an airburst explosion at
an altitude of 37~km, with an energy equivalent to about one kiloton
of TNT explosives. This explains why Figure~\ref{fig:all_obs_TC3_zoom}
and Figure~\ref{fig:all_obs_TC3} also show the regions at altitudes
37~km and 65.4~km, in addition to those on the ground and at
$h=100$~km. The blue line is the nominal ground track. Moreover, the
locations of the recovered meteorites reported in
\cite{shaddad_etal_2010} are displayed, with larger and darker circles
for larger masses. We show with different colours the different impact
regions, according to the displayed legend.  The south displacement of
the smaller meteorites with respect to the predicted ground track is
in agreement with the results of \cite{fjrcdc2017}, where it is argued
that the displacement is likely caused by winds.

\begin{figure}[!ht]
  \centering
  \begin{minipage}{.48\textwidth}
    \includegraphics[width=\textwidth]{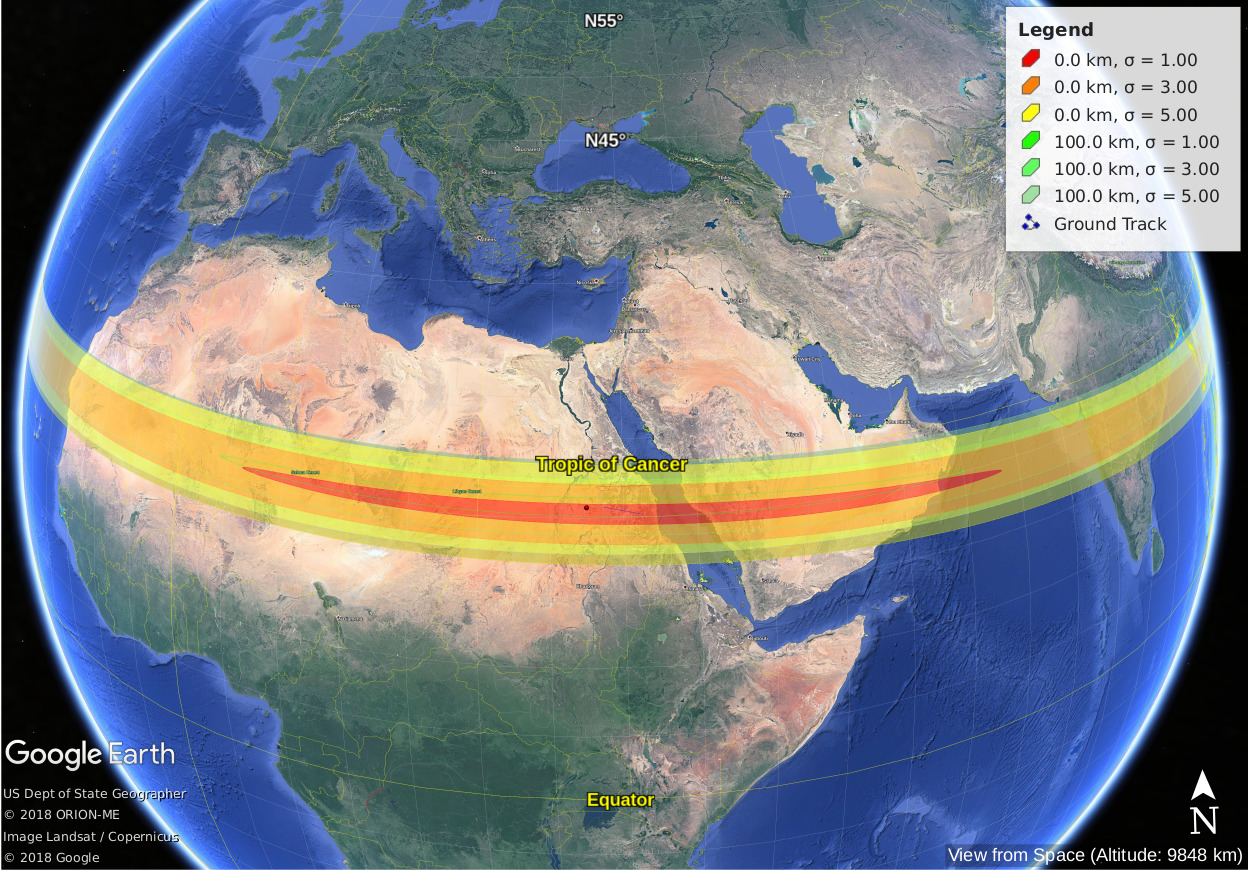}
    \caption{Prediction of the impact region of 2008~TC$_3$ using the
      first 12 observations (almost 18 hours before impact). }
    \label{fig:1red_obs_TC3}
  \end{minipage}
  \hspace{0.2 cm}
  \begin{minipage}{.48\textwidth}
    \includegraphics[width=\textwidth]{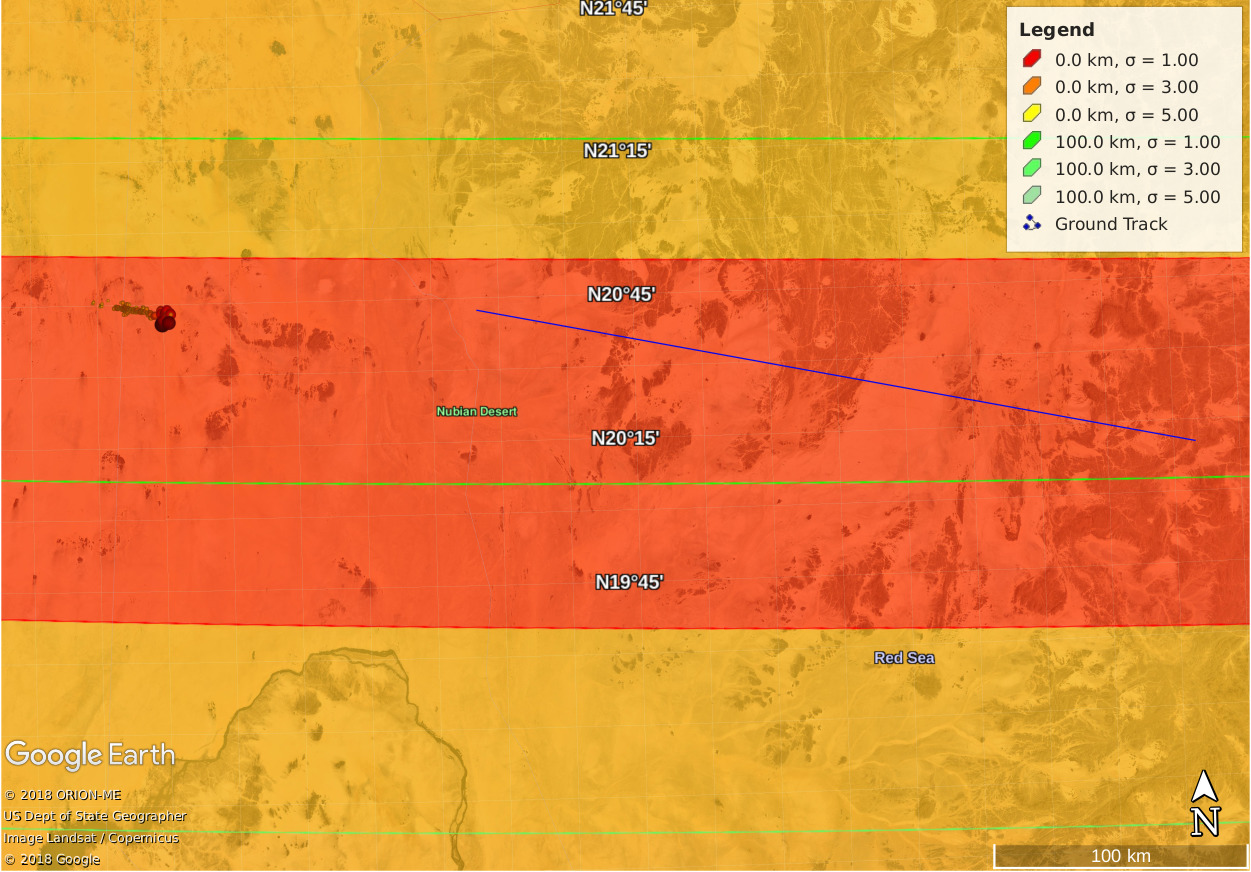}
    \caption{Enlargement of Figure \ref{fig:1red_obs_TC3}, showing the ground track and the locations of the recovered meteorites. }
    \label{fig:1red_obs_TC3_enlarged}
  \end{minipage}
\end{figure}

\begin{figure}[!ht]
  \centering
  \includegraphics[width=0.94\textwidth]{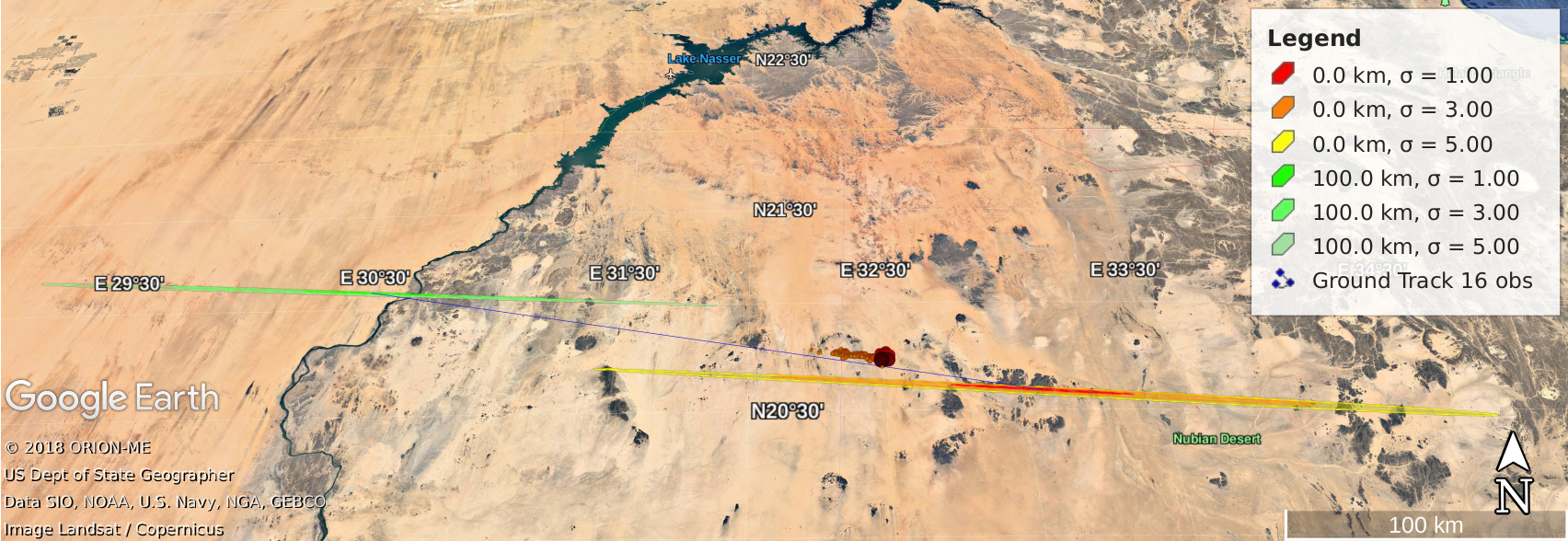}
  \caption{Prediction of the impact region of 2008~TC$_3$ using the
  first 16 observations (about 12 hours before impact).}
  \label{fig:2red_obs_TC3}
\end{figure}

\begin{figure}[!ht]
  \centering
  \includegraphics[width=0.94\textwidth]{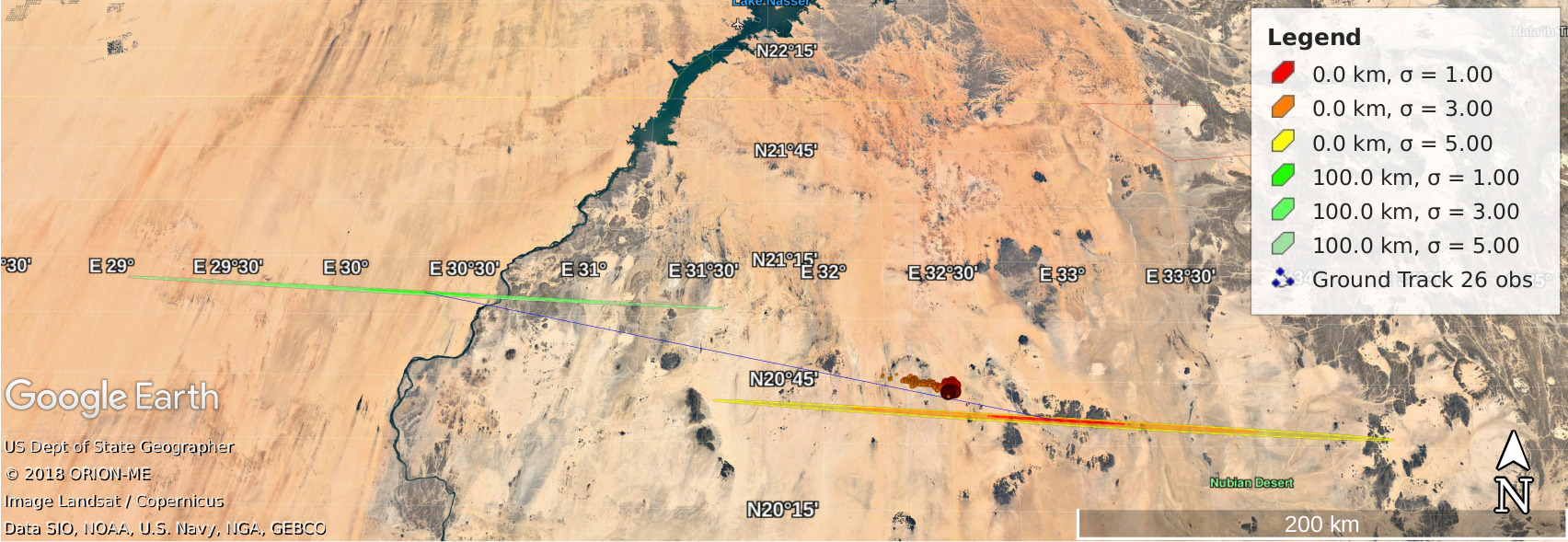}
  \caption{Prediction of the impact region of 2008~TC$_3$ using the
  first 26 observations (about 11 hours and a half before
  impact).}
  \label{fig:3red_obs_TC3}
\end{figure}

In the second test we analysed the evolution of the impact corridor
prediction, using different reduced observational data sets. For any
selected batch of observations we performed the entire set of impact
monitoring computations comprising the orbit determination, the impact
prediction (VI search and characterisation), the IP computation and,
as last step, the computation of the impact corridor.
We selected as input the first 12, 16, and 26 observations,
respectively. Even with so few observations the predicted IP reaches
$1$. For any batch of observations we computed the impact regions at
altitudes 100 and 0~km, for different confidence levels, namely
$\sigma=1,3,5$. A Google Earth 3D visualisation of the impact regions
is reported in Figure~\ref{fig:1red_obs_TC3}, \ref{fig:2red_obs_TC3}
and \ref{fig:3red_obs_TC3}.

In the case with only 12 observations, represented in Figure
\ref{fig:1red_obs_TC3}, the uncertainty is still high, so that the
impact region is very extended and surrounds the entire Earth. In this
figure, the most part of the green regions at 100~km altitudes is not
visible, because they overlap with the regions at 0~km altitude, that
have the priority in the plot. Anyway, all the boundaries are
visible. Moreover the ground track is far from the locations of the
recovered meteorites. Nonetheless they are well inside the 1-$\sigma$
prediction, as it can be seen in the enlargement in Figure
\ref{fig:1red_obs_TC3_enlarged}.

\begin{figure}[!ht]
  \centering
  \includegraphics[width=0.9\textwidth]{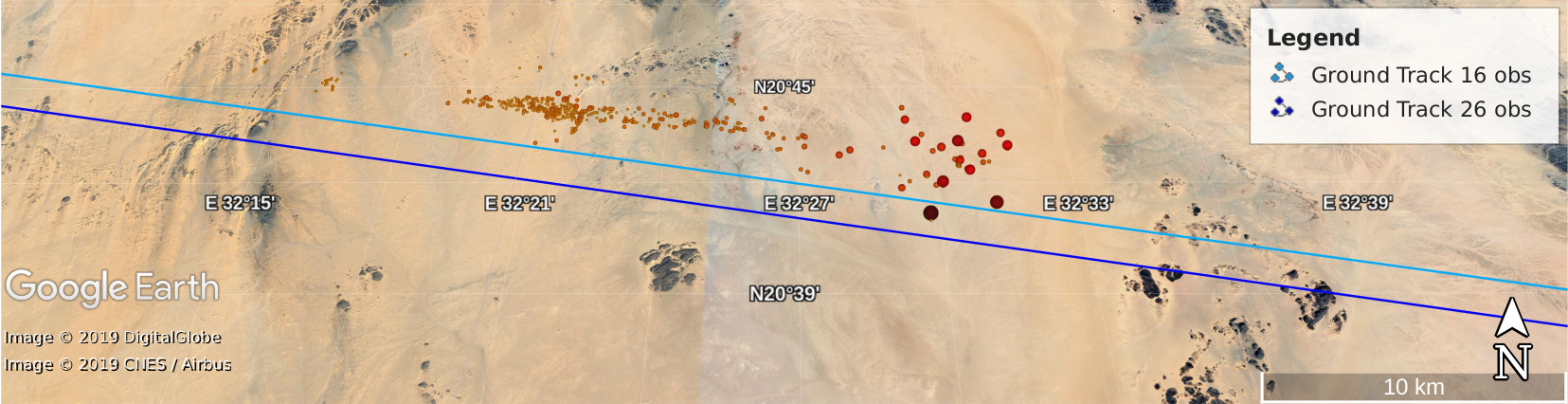}
  \caption{Ground tracks computed with 16 (light blue line) and 26
  (blue line) observations, compared with the locations of the
  recovered 2008~TC$_3$ meteorites (red circles).}
  \label{fig:ground_tracks}
\end{figure}

\begin{figure}[!ht]
  \centering
  \includegraphics[width=0.9\textwidth]{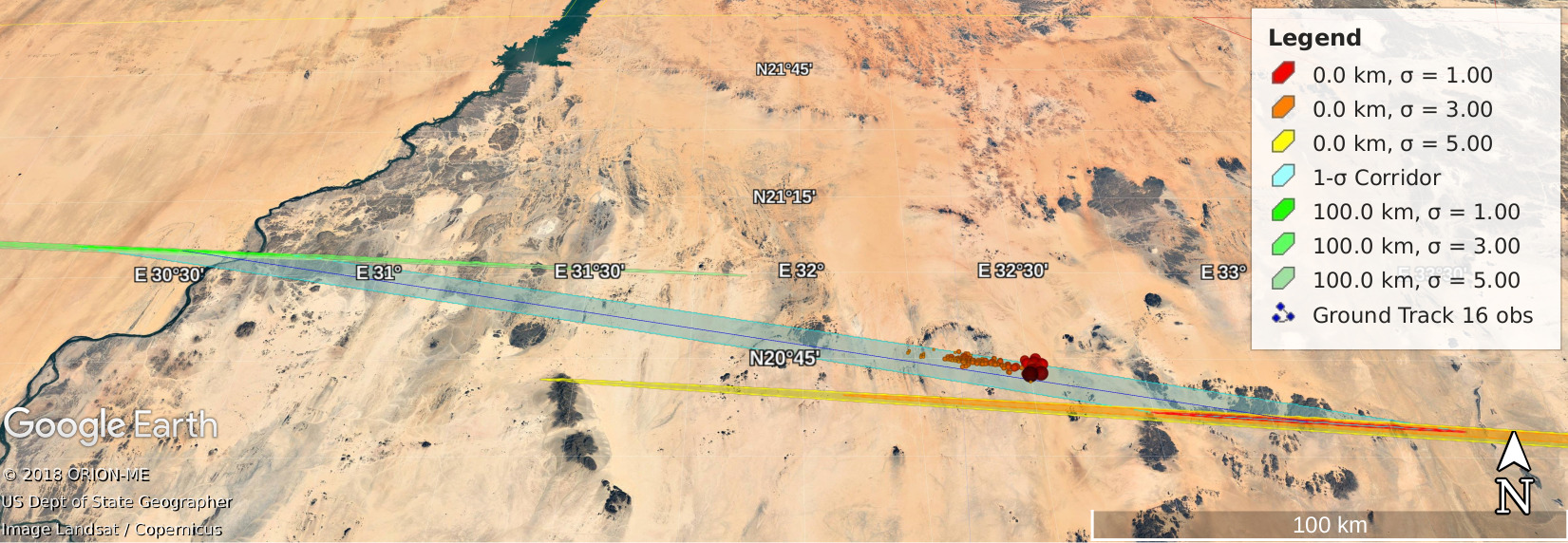} \caption{Approximated
    view of the 1-$\sigma$ impact corridor of 2008~TC$_3$ obtained
    with 16 observations.}
  \label{fig:ic_16obs}
\end{figure}

\begin{figure}[!ht]
  \centering
  \includegraphics[width=0.9\textwidth]{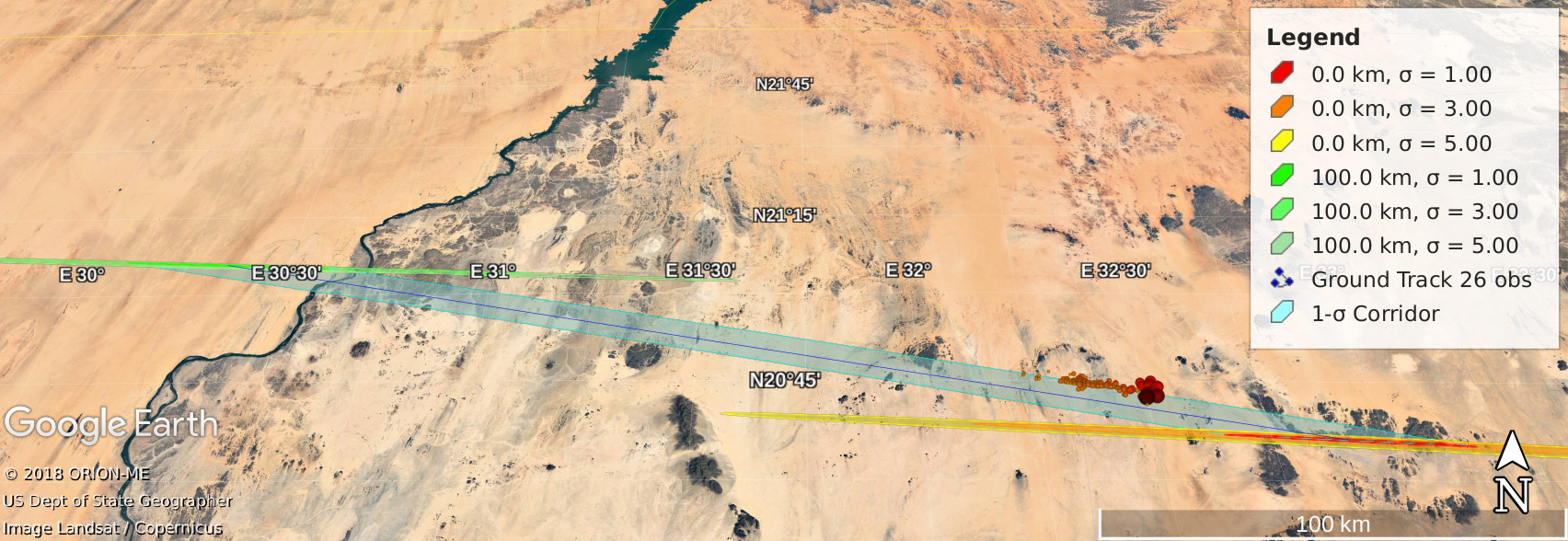}
  \caption{Approximated
    view of the 1-$\sigma$ impact corridor of 2008~TC$_3$ obtained with
    26 observations.}
  \label{fig:ic_26obs}
\end{figure}

With 16 observations, the 1-$\sigma$ impact regions are about 70~km
large. With 26 observations the 1-$\sigma$ impact regions size
decreases to 40~km. For this cases the predicted ground track is near
the locations of the recovered meteorites, even if they does not lie
around it as it would be if the prediction was exact. Indeed, the
ground track corresponding to 16 observations is shifted about half km
southworth with respect to the ideal line crossing the middle of the
meteorite locations. Surprisingly, the ground track corresponding to
26 observations is about 2~km farther toward South (see Figure
\ref{fig:ground_tracks}). This causes a bigger part of the meteorites
impact locations to be placed outside the 1-$\sigma$ impact corridor,
whose approximated plot is obtained joining the semimajor axis
endpoints at 100 and 0~km altitudes, see Figure~\ref{fig:ic_16obs} and
\ref{fig:ic_26obs}.

From this second test it is clear that the impact region prediction,
with whatever method, cannot bring to practical decisions, like for
example the decision about to evacuate the region interested by a
possible impact or even the selection of the place where to organise a
mission for the recovery of meteorites after the impact occurred, when
the uncertainty is too high. On the other hand, only by performing the
prediction we have the precise measure of the size of the impact
region on ground. In this particular test, only when at least 16
observations are available, we obtain a restricted impact area on
ground.

\subsection{2004~MN$_4$ - Apophis: Impact Prediction with 2004 Data}
\label{sec:apophis}

%
Apophis was first observed on June 19$^{\rm th}$, 2004 by R. Tucker,
D. Tholen and F. Bernardi from Kitt Peak over two consecutive
nights. It was not observed for the following six months and it was
recovered by chance on December 18$^{\rm th}$ by G. Garradd, who
observed it for three consecutive nights. The object was recognised to
be the one observed in June, with provisional designation 2004~MN$_4$,
and the MPC released the new data of the recovered asteroid on
December 20$^{\rm th}$. With these data, both the \clomon-2 a Sentry
IM systems found the possibility of an impact in 2029.
Anyway, due to some problems in the processing of the raw images, the
reduced measurements of June were spoiled and consequently the fit
showed very high residuals corresponding to those data. Consequently, the
prediction of an impact was not trusted by the scientists, who asked for new
data.

Already on December 23$^{\rm rd}$, the availability of new
observations and of the remeasurements of the bad June observations
allowed the IM systems to provide a more reliable result. At this
point, both \clomon-2 and Sentry gave a VI in 2029 with Torino Scale
\cite{morrison_etal2004,binzel2000} value 2 and Palermo Scale
\cite{chesley_etal2002} greater than zero. The agreement between the
systems was good and, even if the distribution and the quality of the
observations was not optimal, the result was published. A note was
added in the Risk Page, clarifying that this result was subject to
change when new measurements would be available.

In the following days, after the addition of new incoming
observations, the impact prediction continued to be confirmed and the
impact probability grew to its maximum on December 27$^{\rm th}$, when
the IM systems gave $IP = 0.027$ (1 in 38).
A case like this had never happen before and, in addition to the near
and high risk of impact, it presented new dynamical features which the
\clomon-2 software was not ready to deal with. It was a challenge for
the NEODyS team to face the difficulties arising hour by hour,
changing all the parts of the software that were not working properly
in the minimum possible time.

The situation changed during the afternoon of December 27$^{\rm th}$ with
the issue of 4 new special MPECs by the MPC. The last one, the
MPEC-Y70, contained pre-discovery observations going back to March
2004. With those observations, the possibility of an impact in 2029
was ruled out during the night between December 27$^{\rm th}$ and
28$^{\rm th}$, when the processing of the new data set was completed.

\begin{figure}[!ht]
  \centering
  \includegraphics[width=0.8\textwidth]{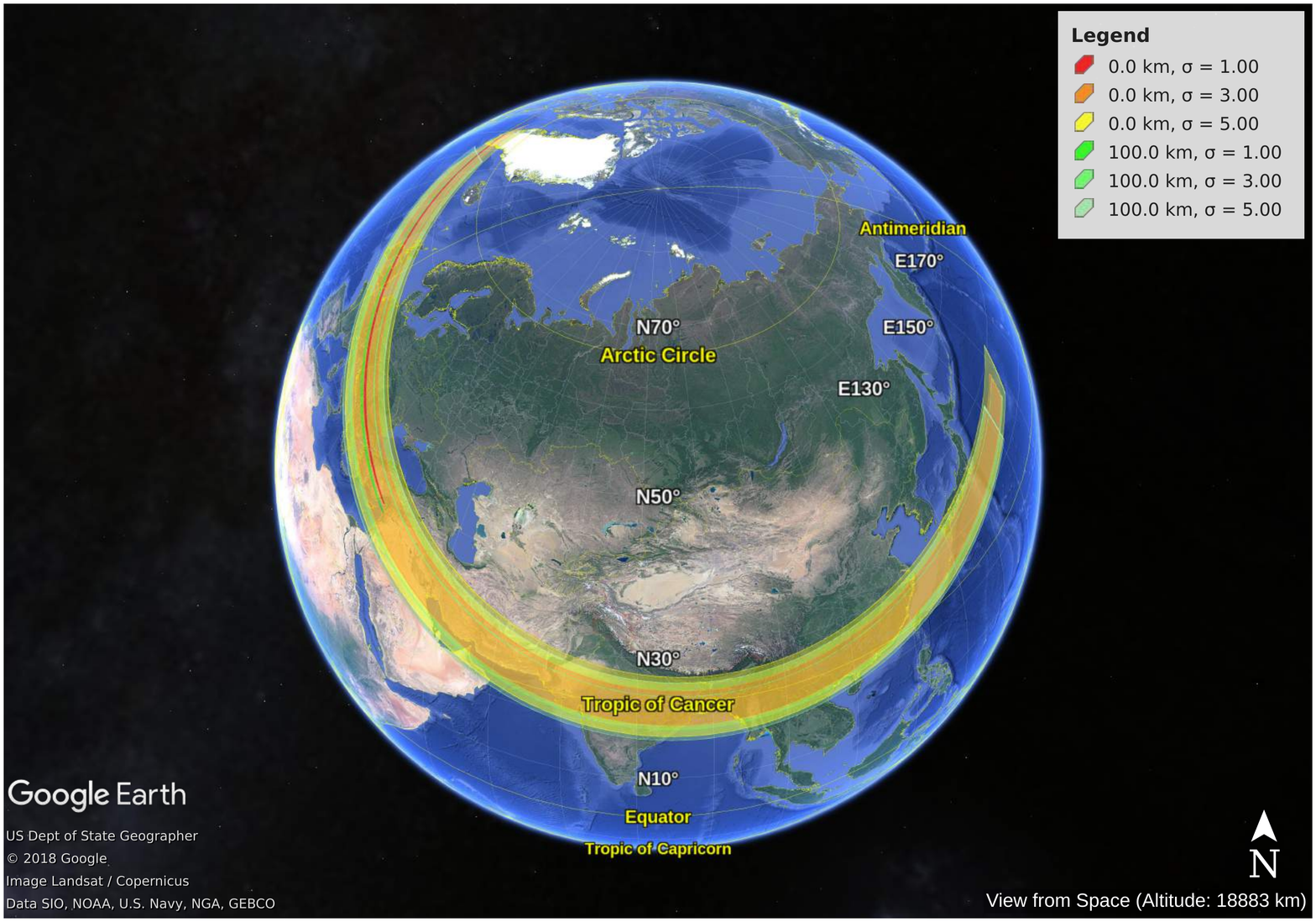}
  \\
  \includegraphics[width=0.5\textwidth]{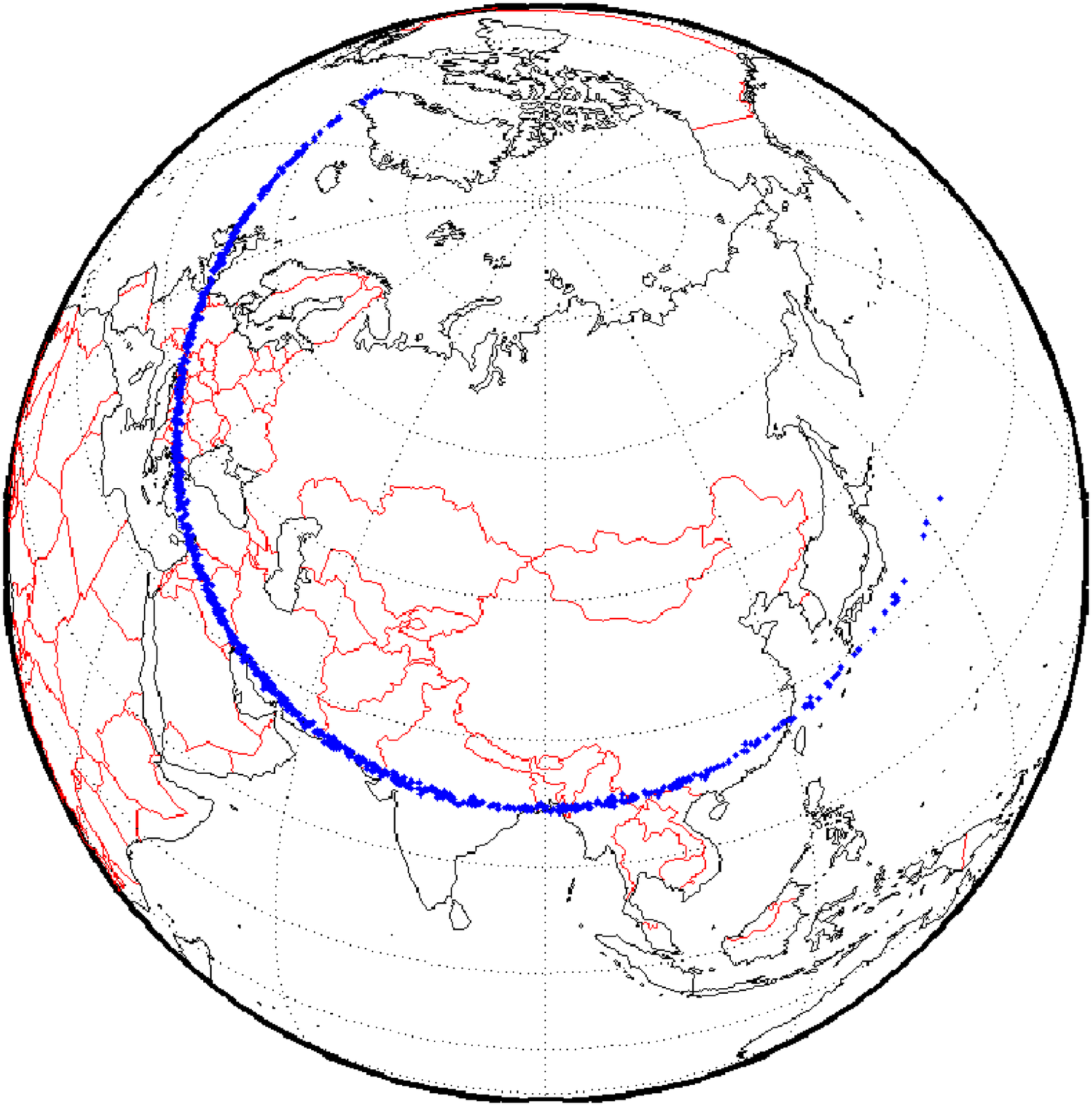}
  \caption{\emph{Top panel}. Google Earth 3D visualisation of the
  semilinear prediction of the 2029 impact regions of Apophis, using
  the observations available on December 27$^{\rm th}$, 2004 (MPEC
  2004-Y69, 18:40 UT). \emph{Bottom panel}. Monte Carlo prediction
  of the 2029 possible impact locations on ground of Apophis, using
  the same observational dataset of the above figure (JPL, private
  communication).}
  \label{fig:2004MN4}
\end{figure}

The details about the Apophis discovery and the story of the
\emph{Apophis crisis} following the prediction of an impact in 2029
can be found in \cite{sansaturio-arratia2008}.
We recovered the situation for Apophis corresponding to December
27$^{\rm th}$, 2004. The set of observations taken is the one of MPEC
2004-Y69. This set corresponds to the situation just before the
availability of pre-discovery observations.
Using OrbFit version 5.0, we computed a full least squares solution
and we performed the impact monitoring with the options specified at
the beginning of Section \ref{sec:tests}.
The result of the IM gave a VI with impact on April 13$^{\rm th}$, 2029 with
probability $0.0242$ (1 in 41). The semilinear impact corridor was computed
starting from this result, see top panel of Figure
\ref{fig:2004MN4}. The bottom panel of this figure shows the impact
region computed with the same observational dataset by the JPL team
using a Monte Carlo method\footnote{Private communication by Steven
  R. Chesley.}.
As it can be seen from the figures, there is a good agreement between
the two independent predictions.

\subsection{Impact Corridor for 2014~AA}
\label{sec:2014AA}

Asteroid 2014~AA was discovered by R. Kowalski of the Catalina Sky
Survey on January 1$^{\rm st}$, 2014 at 06:18 UTC. The asteroid
impacted the Earth on January 2$^{\rm nd}$, 2014 at about 3 UTC, less
than 21 hours after the first detection. The situation was similar to
2008~TC$_3$, but in this case very few observations were available
before the impact, with a total of 7 optical measurements. Even with
these few observations, the computed $IP$ was equal to $1$.  The
infrasonic airwaves produced by the 2014~AA atmospheric impact were
detected by the infrasound component of the International Monitoring
System operated by the Comprehensive Nuclear Test Ban Treaty
Organization, as reported in \cite{fcbc2016}.

\begin{figure}[!ht]
  \centering
  \includegraphics[width=0.6\textwidth]{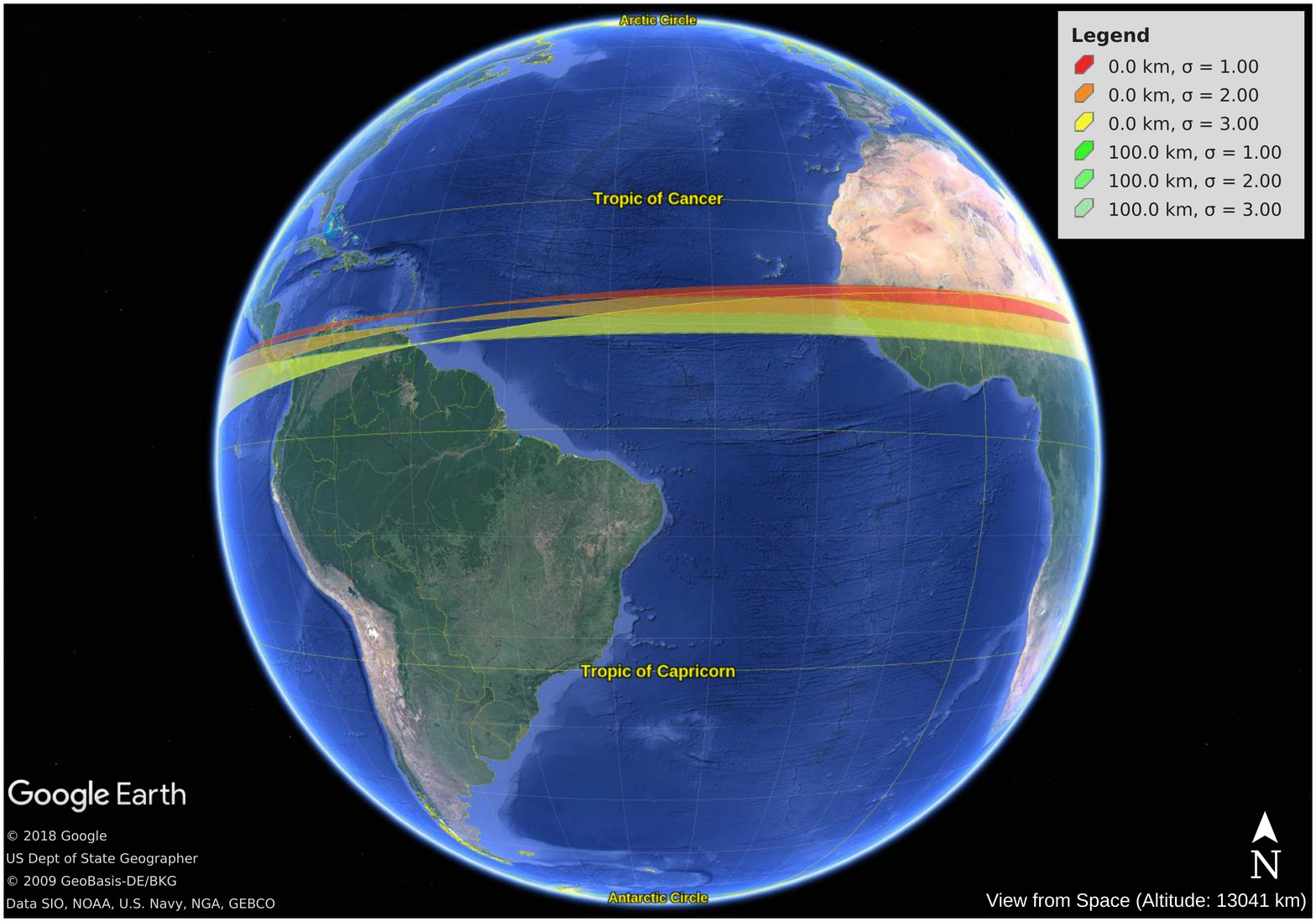}
  \caption{Google Earth 3D visualisation of 2014~AA impact regions for
  $\sigma$= 1, 2, 3 on ground and for 100~km altitude.}
  \label{fig:2014AA}
\end{figure}

We have applied the semilinear method using the impacting nominal
solution as VI representative.  In Figure~\ref{fig:2014AA}, the
semilinear impact regions on ground and at 100~km altitude are
shown. The regions at the two altitudes almost overlap. This case is
remarkable, because it is extremely non-linear, with the regions
twisting on themselves. It follows that the coloured regions of the
figure do not correspond to the interior of the regions delimited by
the illustrated boundaries. The impact regions actually extend outside
the drawn boundaries (the Jordan curve theorem does not apply because
of the self-intersection of the non-linear image of the ellipse). This
is confirmed by the fact that, in the vicinity of the torsion, the
boundaries with lower $\sigma$ extends outside the ones with higher
$\sigma$.

\begin{figure}[!ht]
  \centering
  \includegraphics[width=0.7\textwidth]{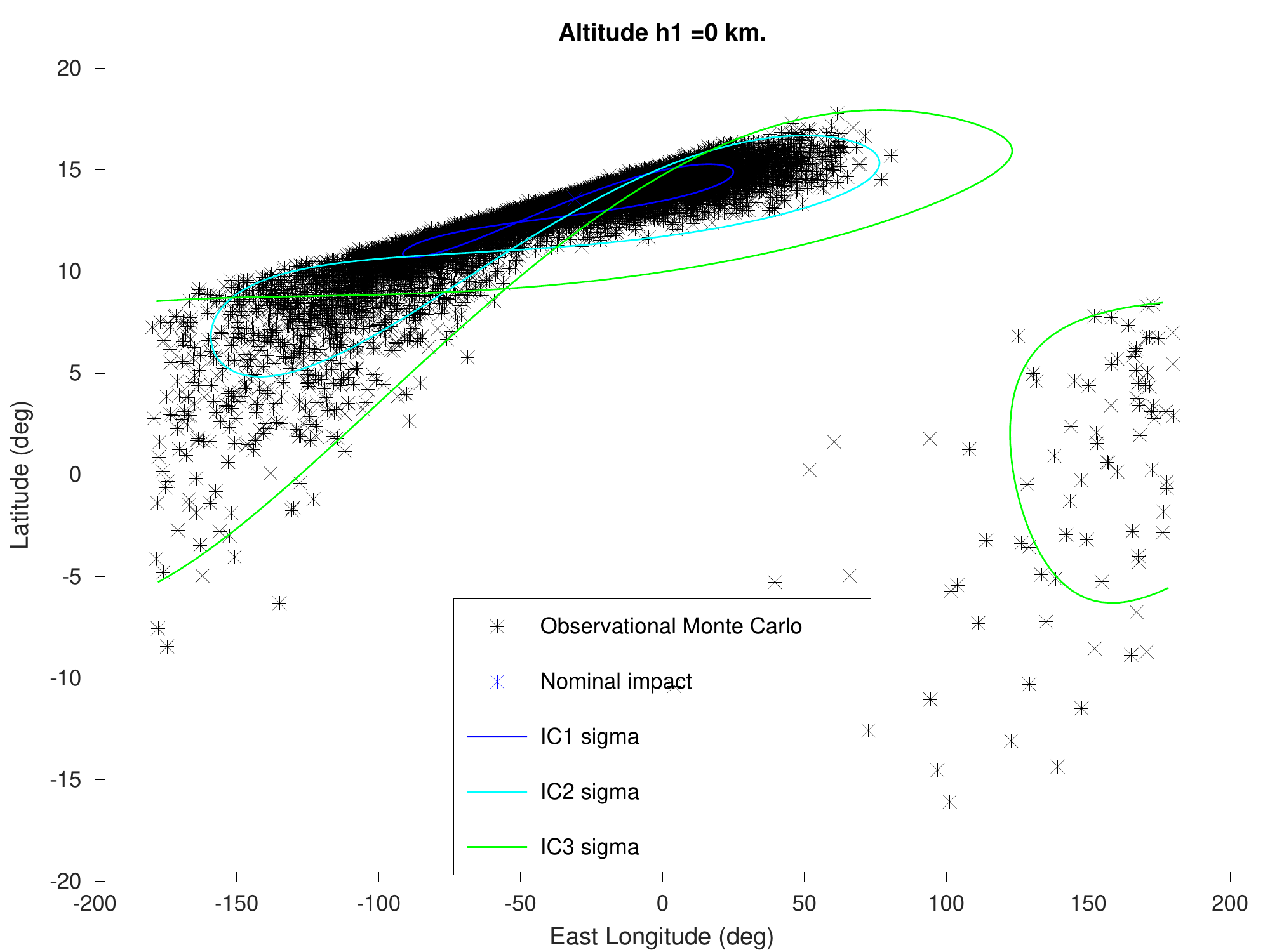}
  \caption{Comparison of the Monte Carlo prediction and the 2014~AA semilinear impact region boundaries on ground for $\sigma$= 1, 2, 3.}
  \label{fig:2014AA_mc}
\end{figure}

The region obtained by applying an observational Monte Carlo method is
consistent with the boundaries obtained with the semilinear approach
(see Figure~\ref{fig:2014AA_mc}), which confirms that the semilinear
prediction is correct. The problem is that the boundaries given by the
semilinear method in this case does not provide complete information
on the impact region extension.
In conclusion, for cases like this one (the only one encountered so
far, considering also other tests not reported here) the information
provided by the semilinear boundary prediction is incomplete and can
be misleading for a non-specialist audience.

\subsection{Recent imminent impactors 2018~LA and 2019~MO}
\label{sec:2018LA_2019MO}

Object 2018~LA was a small (2-3 metres in diameter) Apollo near-Earth
asteroid. It was discovered 8 hours before impact by the Mt. Lemmon
Observatory of the Catalina Sky Survey. The impact occurred on June
2$^{\rm nd}$, 2018 at 16:44 UTC (18:44 local time) in Botswana.

\begin{figure}[!ht]
  \centering
  \includegraphics[width=0.9\textwidth]{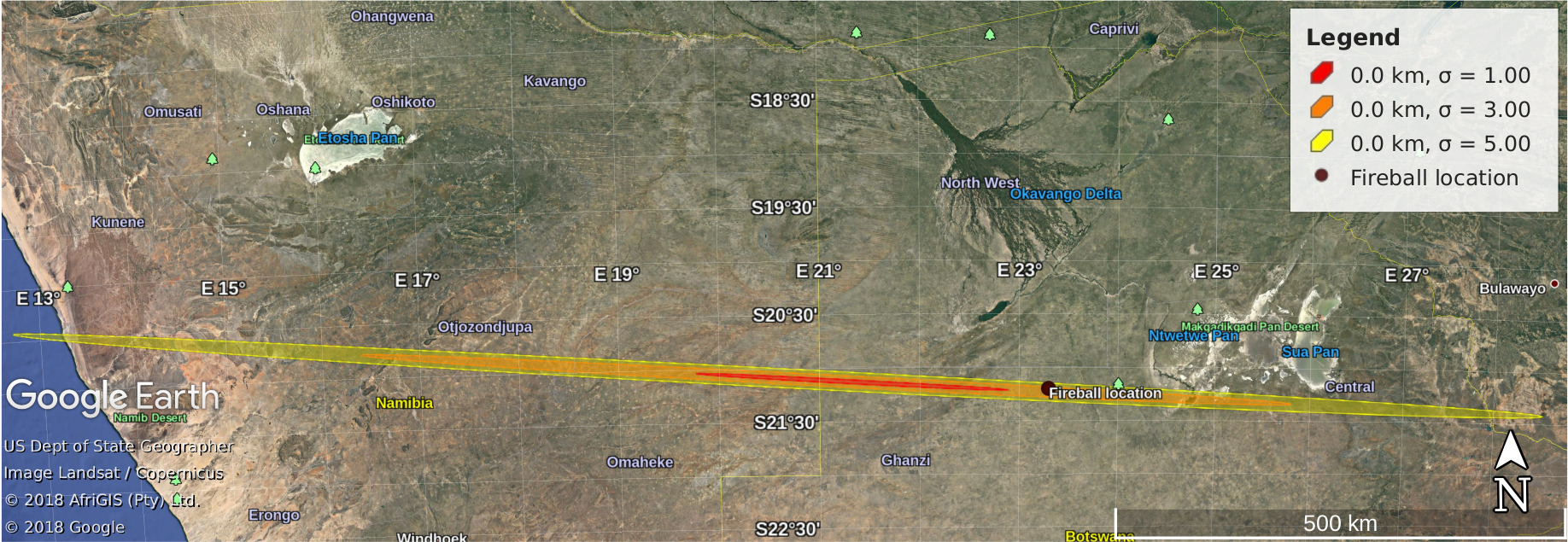}
  \caption{Google Earth 3D visualisation of 2018~LA impact regions on ground for
  $\sigma$= 1, 3, 5.
  The fireball location at 28.7~km altitude is also displayed. }
  \label{fig:2018LA}
\end{figure}

The orbit determination and IM computations were performed using all
the available 14 observations and gave $IP = 1$. The semilinear
algorithm was used to compute the impact regions on ground
corresponding to $\sigma = 1,3,5$, which are shown in
Figure~\ref{fig:2018LA} together with the firewall location, whose
coordinates are extracted from the JPL fireballs web-page and reported
in Table~\ref{tab:fireball_rep}. As it can be seen from the figure,
the fireball location is inside the 3-$\sigma$ prediction.

Object 2019~MO was discovered from the ATLAS Mauna Loa observatory on
June 22$^{\rm nd}$, 2019 at 9:49 UTC. This object was a small (4-6 metres
in diameter) Apollo near-Earth asteroid. It was discovered less than
12 hours before impact, which occurred on June 22$^{\rm nd}$, 2019 at
21:25 UTC, between Jamaica and south American coast.

At the beginning, only four observations from the Mauna Loa site were
available and the object was in the NEO Confirmation Page of the MPC,
which contains unconfirmed objects.
The Italian
NEOScan\footnote{\url{https://newton.spacedys.com/neodys2/NEOScan/}}
\cite{spoto_etal2018} and the JPL
Scout\footnote{\url{https://cneos.jpl.nasa.gov/scout/\#/}}
\cite{farnocchia_etal2015} systems routinely scan this page with the
goal to find imminent impactors.

The object 2019~MO was indeed found to be an impactor by these
systems, but with a low score, so that follow-up observations were not
started promptly. Additional observations from the Pan-STARRS2 images
were recovered only after the impact, on June 25$^{\rm th}$. After this
precovery, 7 observations were available and the fit was good enough
to remove the object from the NEOCP and release an MPEC, but the
object had already impacted the Earth.  Anyway, post-impact
computations were performed by the IM systems, giving an IP
practically equal to 1 ($IP = 0.994$ by \clomon-2).

\begin{figure}[!ht]
  \centering
  \includegraphics[width=0.9\textwidth]{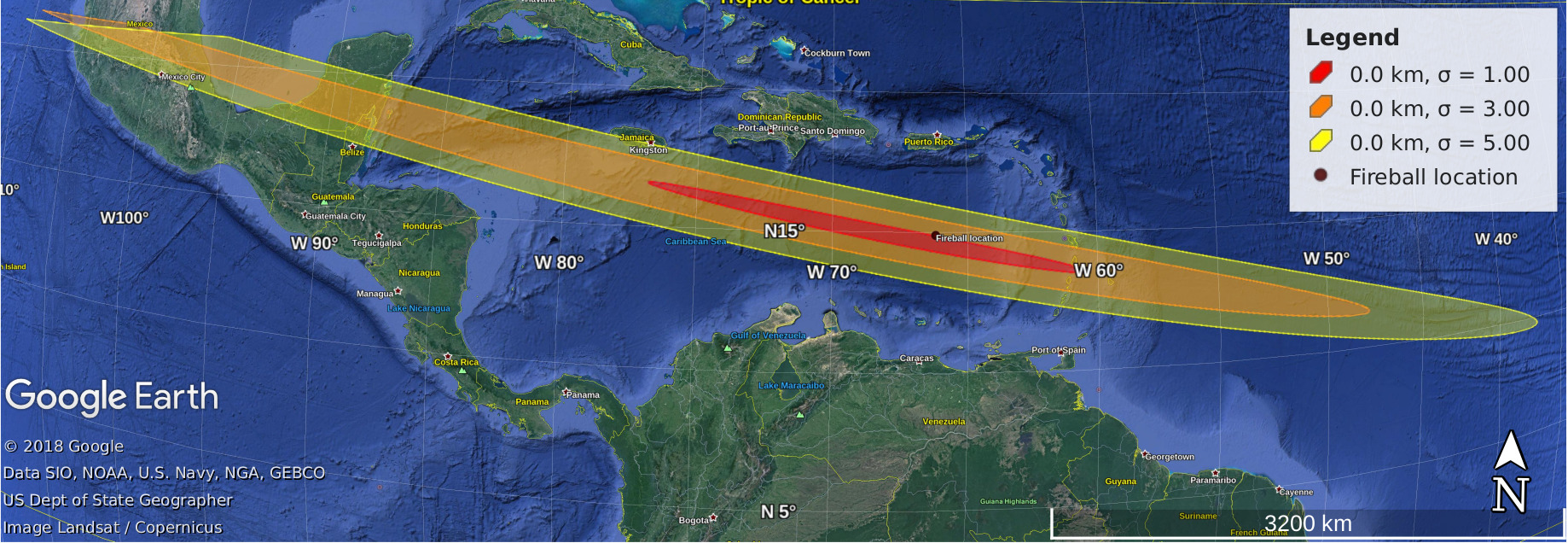}
  \caption{Google
  Earth 3D visualisation of 2019~MO impact regions on ground for
  $\sigma$= 1, 3, 5 and the fireball location at 25~km
  altitude.}  \label{fig:2019MO}
\end{figure}

We computed the impact corridor for this object using the solution
obtained with the entire set of 7 observations. The result is an
impact region almost centred on the real impact location, which is
known, because the fireball corresponding to the 2019~MO impact was
observed. The semilinear impact region on ground and the fireball
location at 25~km altitude are shown in Figure~\ref{fig:2019MO}. The
fireball data reported on the fireballs page of the JPL web-site are
shown in Table~\ref{tab:fireball_rep}.

\begin{table}[!ht]
  \caption{Fireball reports corresponding to 2018~LA and 2019~MO
    impacts, extracted from
    \url{https://cneos.jpl.nasa.gov/fireballs/}.}
  \label{tab:fireball_rep}
  \centering
  \begin{tabular}{p{0.14\linewidth}p{0.1\linewidth}p{0.1\linewidth}p{0.1\linewidth}p{0.1\linewidth}p{0.12\linewidth}p{0.1\linewidth}}
    \toprule
    \textbf{Peak Brightness Date/Time (UT)} &
    \textbf{Latitude (deg)} &
    \textbf{Longitude (deg)} &
    \textbf{Altitude (km)} &
    \textbf{Velocity (km/s)} &
    \textbf{Total Radiated Energy (J)} &
    \textbf{Calculated Total Impact Energy (kt)}\\
    \toprule
    2019-06-22 21:25:48	& 14.9N	& 66.2W & 25.0 & 14.9 & 294.7e10 & 6 \\
    \midrule
    2018-06-02 16:44:12	& 21.2S	& 23.3E	& 28.7 & 16.9 & 37.5e10	& 0.98\\
    \bottomrule
  \end{tabular}
\end{table}

As a final remark, we note that this kind of impact events occur quite
frequently, but they are difficult to be predicted, because the
objects are difficult to be observed before the impact. This case was
the fourth after 2008~TC$_3$, 2014~AA and 2018~LA to be discovered
just before the impact and having enough observations for a reliable
orbit determination. Apart from 2014~AA, for all the other cases the
semilinear algorithm succeeded in giving reliable predictions of the
impact regions, even with few observations like 2019~MO.

%

\section{Conclusions and future work}
\label{sec:conclusions}
The semilinear method succeeds in providing the boundary of the impact
region on ground, with a comparatively smaller number of propagations
with respect to Monte Carlo approaches. Indeed, it samples a
1-dimensional curve instead of a region in the 6-dimensional orbital
elements space.

It is possible to compute the impact probability associated with each
boundary, absolute or conditional. For $IP=1$ it coincides with the
probability of the orbital solution and is given by the level
$\sigma$ of the considered impact region.

The algorithm has been tested using the real observations of the past
impacted objects 2008~TC$_3$, 2014~AA, 2018~LA and 2019~MO. It has
been applied also to the asteroid Apophis, but using only the
observations available on December 27$^{\rm th}$, 2004, before that
pre-discovery observations were found. This situation corresponds to a
possibility of impact in 2029 with probability of about 2.4\%, as
computed by the last version of OrbFit, version 5.0. For 2008~TC$_3$
and Apophis the predicted impact regions on ground are in good
agreement with the Monte Carlo predictions by the JPL-NASA system.
The semilinear prediction is especially good for 2008~TC$_3$, for
which the predicted thin impact corridor along the ground track passes
through the region of recovered meteorites. For 2018~LA and 2019~MO,
the predicted semilinear impact regions contain the locations of the
observed fireballs, even if very few observations are available.  Only
the case of 2014~AA reveals some limitations of the method. In this
case the non-linearity causes the propagated uncertainty ellipse to
twist on itself, so that the drawn boundary of the impact region does
not encompass the inner points, and it provides incomplete and
misleading information.

The performance of the implemented algorithm is good. It can be
further improved using parallelisation. In particular, parallel
computing is useful for special cases with low $IP$, near to the
threshold of $10^{-3}$, with far impact time in the future and for
which the implemented optimisation procedure cannot be applied, for
example if there are multiple VIs with the same impact date. These
cases correspond to a high level of non-linearity, but the semilinear
prediction reveals to be still reliable. The only case in which the
information provided could be misleading is when the effect of
non-linearity is so high to cause a twist of the propagated
uncertainty region, as it happens for the object 2014~AA.
The best results in terms both of performance and accuracy are
obtained for high impact probabilities and impact time near to the
current date.  It follows that the semilinear method is particularly
useful for imminent impactors, as shown by the application of the
method for the cases of the past impacted objects 2008~TC$_3$, 2018~LA
and 2019~MO, for which the software takes between 30 and 50 seconds of
runtime for each single impact region, with fixed altitude and $\sigma$,
without the need of parallelisation.

The current software does not perform the computation of the IP
associated with the impact regions.
The current algorithm can be refined modifying the selection of the VI
representative, in order to obtain a representative located as near as
possible to the centre of the VI. This improvement is not
straightforward. It requires a conversion of the LOV parameter to the
corresponding physical distance along the LOV. Moreover, the case of a
VI expanding off-LOV needs special attention.

A higher level improvement can be obtained combining together the
semilinear method and the projection of the non-linear LOV on ground to
obtain a more accurate prediction of the impact region.

\section*{Acknowledgements}
We dedicate this work to the memory of Prof. Andrea Milani
Comparetti. His contribution for the development of the method for the
impact corridor computation was fundamental and he would have written
this paper with us if he was alive. We owe a lot to his teachings.

Special thanks are reserved to Prof. Giovanni Valsecchi for his help
in recovering the historical situation of 2004~MN$_4$, namely
(99942)~Apophis, corresponding to the collision scenario evolution
during Christmas 2004.

\bibliographystyle{elsarticle-harv} \bibliography{impcor}

\end{document}